\documentstyle[12pt,amssymb]{article}

\def\fnote#1#2{\begingroup\def\thefootnote{#1}\footnote{#2}\addtocounter{footnote}{-1}\endgroup}

\def\inbar{\vrule height1.5ex width.4pt depth0pt}
\def\IB{\relax{\rm I\kern-.18em B}}
\def\IC{\relax\,\hbox{$\inbar\kern-.3em{\rm C}$}}
\def\ID{\relax{\rm I\kern-.18em D}}
\def\IE{\relax{\rm I\kern-.18em E}}
\def\IF{\relax{\rm I\kern-.18em F}}
\def\IG{\relax\,\hbox{$\inbar\kern-.3em{\rm G}$}}
\def\IH{\relax{\rm I\kern-.18em H}}
\def\II{\relax{\rm I\kern-.18em I}}
\def\IK{\relax{\rm I\kern-.18em K}}
\def\IL{\relax{\rm I\kern-.18em L}}
\def\IM{\relax{\rm I\kern-.18em M}}
\def\IN{\relax{\rm I\kern-.18em N}}
\def\IO{\relax\,\hbox{$\inbar\kern-.3em{\rm O}$}}
\def\IP{\relax{\rm I\kern-.18em P}}
\def\IQ{\relax\,\hbox{$\inbar\kern-.3em{\rm Q}$}}
\def\IR{\relax{\rm I\kern-.18em R}}
\def\IT{\relax{\rm I\kern-.18em T}}
\def\ZZ{\relax{\sf Z\kern-.4em Z}}

\def\a{\alpha}   \def\b{\beta}    \def\g{\gamma}  
      
  \def\om{\omega}  \def\Om{\Omega} \def\si{\sigma}

\def\cA{{\cal A}} \def\cB{{\cal B}}
   
  \def\cI{{\cal I}} 
   
\def\cO{{\cal O}} \def\cP{{\cal P}}

\def\afrak{{\mathfrak a}} \def\bfrak{{\mathfrak b}} 
\def\ffrak{{\mathfrak f}}
\def\pfrak{{\mathfrak p}} \def\qfrak{{\mathfrak q}}
\def\Afrak{{\mathfrak A}} 
\def\Ifrak{{\mathfrak I}} \def\Pfrak{{\mathfrak P}}

\def\mathA{{\mathbb A}}   
\def\mathC{{\mathbb C}}   
\def\mathN{{\mathbb N}}   \def\mathQ{{\mathbb Q}}
\def\mathR{{\mathbb R}}   \def\mathZ{{\mathbb Z}}

\def\bx{{\bar x}}    \def\by{{\bar y}}

 \def\bF{{\bar F}}

  \def\bchi{{\bar \chi}}

\def\bsi{\bar \sigma}

\def\Fhat{{\hat F}}   

\def\Phihat{{\hat \Phi}} \def\vphihat{{\hat \vphi}}

 \def\ZZmum{{\ZZ[\mu_m]}}

\def\fnote#1#2{\begingroup\def\thefootnote{#1}\footnote{#2}\addtocounter
{footnote}{-1}\endgroup}
\def\beq{\begin{equation}}
\def\eeq{\end{equation}}
\def\bea{\begin{eqnarray}}
\def\eea{\end{eqnarray}}

\def\lleq#1{\label{#1}\eeq}
\let\nn=\nonumber
\def\tabroom{\hbox to0pt{\phantom{\Huge A}\hss}}
\def\notin{\ \hbox{{$\in$}\kern-.51em\hbox{/}}}

\def\lra{\longrightarrow}

\def\ra{{\rightarrow}}

\def\vphi{\varphi}

  \def\E1Fq{E_1/\IF_q}

\def\IFq{{\IF_q}}

 \def\rmH{{\rm H}} 
\def\rmN{{\rm N}}

\def\rmab{{\rm ab}}
\def\rmdR{{\rm dR}}     
\def\rmdet{{\rm det}}   \def\rmdim{{\rm dim}}
\def\rmgcd{{\rm gcd}}   
\def\rmord{{\rm ord}}   \def\rmmod{{\rm mod}}
 \def\rmtor{{\rm tor}}

\def\rmAut{{\rm Aut}}  
\def\rmEnd{{\rm End}}
\def\rmGal{{\rm Gal}}  \def\rmHom{{\rm Hom}}
\def\rmHW{{\rm HW}}   \def\rmIso{{\rm Iso}}

\def\notdiv{{\relax{~|\kern-.35em /~}}}

  \def\E1Fq{E_1/\IF_q} 
  
\def\XFp{{X/\IF_p}}  \def\XFpr{{X/\IF_{p^r}}}
\begin{document}
\parindent=0pt
\hfill {\bf NSF-KITP-03-112}

\vskip 0.5truein

\centerline{\large {\bf COMPLEX MULTIPLICATION OF EXACTLY
SOLVABLE}} \vskip .1truein \centerline{\large {\bf CALABI-YAU
VARIETIES}}

\vskip 0.4truein

\centerline{\sc Monika Lynker$^1$\fnote{\diamond}{email:
mlynker@iusb.edu}, Rolf Schimmrigk$^2$\fnote{\dagger}{email:
netahu@yahoo.com} and Steven Stewart$^2$\fnote{\sharp}{email:
juliusbox@hotmail.com}}

\vskip .3truein

\centerline{\it $^1$Indiana University South Bend}\vskip .05truein
\centerline{\it 1700 Mishawaka Avenue, South Bend, IN 43144}

\vskip .2truein

\centerline{\it $^2$Kennesaw State University} \vskip .05truein
\centerline{\it 1000 Chastain Rd, Kennesaw, GA 30144}

\vskip .5truein

\baselineskip=19pt

\centerline{\bf ABSTRACT:} \vskip .2truein We propose a conceptual
framework that leads to an abstract characterization for the exact
solvability of Calabi-Yau varieties in terms of abelian varieties
with complex multiplication. The abelian manifolds are derived
from the cohomology of the Calabi-Yau manifold, and the conformal
field theoretic quantities of the underlying string emerge from
the number theoretic structure induced on the varieties by the
complex multiplication symmetry. The geometric structure that
provides a conceptual interpretation of the relation between
geometry and the conformal field theory is discrete, and turns out
to be given by the torsion points on the abelian varieties.

\vfill

{\sc PACS Numbers and Keywords:} \hfill \break Math:  11G25
Varieties over finite fields; 11G40 L-functions; 14G10  Zeta
functions; 14G40 Arithmetic Varieties \hfill \break Phys: 11.25.-w
Fundamental strings; 11.25.Hf Conformal Field Theory; 11.25.Mj
Compactification

\renewcommand\thepage{}
\newpage
\pagenumbering{arabic}

\baselineskip=16.5pt

\tableofcontents

\vfill \eject

%ar: 22
%st: 20.7
%bh: 22.2

\baselineskip=18.5pt
\parskip=.21truein
\parindent=0pt

\section{Introduction}

Arithmetic properties of exactly solvable Calabi-Yau varieties
encode string theoretic information of their underlying conformal
field theory. Results in this direction address the issue of an
intrinsic geometric description of the spectrum of the conformal
field theory, and a geometric derivation of the characters of the
partition function. The computations that have been performed so
far depend on the explicit computation of the Hasse-Weil
L-function of Fermat varieties, or more generally Brieskorn-Pham
type spaces. The special feature of these manifolds, first
observed by Weil \cite{w49, w52} about fifty years ago, is that
the cohomological L-function can be expressed in terms of number
theoretic L-functions, defined by special kinds of so-called
Gr\"o\ss encharaktere, or algebraic Hecke characters. Weil's
analysis of Fermat type L-functions in terms of Jacobi-sum
Gr\"o\ss encharaktere was generalized by Yui to the class of
Brieskorn-Pham L-functions \cite{y01}. It was shown in \cite{rs01}
that the algebraic number field that emerges from the Hasse-Weil
L-function of an exactly solvable Calabi-Yau variety leads to the
fusion field of the underlying conformal field theory and thereby
to the quantum dimensions. It was further proven in \cite{su02}
that the modular form defined by the Mellin transform of the
Hasse-Weil L-function of the Fermat torus arises from the
characters of the underlying conformal field theory. This
establishes a connection between algebraic varieties and Kac-Moody
algebras via their modular properties.

The basic ingredient of the investigations described in Refs.
\cite{rs01,su02} is the Hasse-Weil L-function, an object which
collects information of the variety at all prime numbers,
therefore providing a 'global' quantity that is associated to
Calabi-Yau varieties. The number theoretic interpretation which
leads to the physical results proceeded in a somewhat experimental
way, by observing the appearance of Jacobi-sum characters in
\cite{rs01}, and that of affine theta functions in \cite{su02}.
This leaves open the question whether these results depend on the
special nature of the varieties under consideration, or whether it
is possible to identify an underlying conceptual framework that
explains the emergence of conformal field theoretic quantities
from the discrete structure of the Calabi-Yau variety. It is this
problem which we address in the present paper.

In principle, the physical question raised translates into a
simply stated mathematical problem: provide a theorem that states
the conditions under which the geometric Hasse-Weil L-function
decomposes into a product of number theoretic L-functions. If such
a statement were known one could ask whether the class of
varieties that satisfies the stated conditions can be used to
derive conformal field theoretic results, e.g. in the spirit of
the results of \cite{rs01,su02}. It turns out that this question
is very difficult. In dimension one it basically is the
Shimura-Taniyama conjecture, which has been recently proven in
full generality by work \cite{bcdt01} that extends the results by
Wiles and Taylor in the semistable case \cite{w95}.

In higher dimensions much less is known, and the problem is often
summarized as the Langlands program, a set of conjectures, which
might be paraphrased as the hope that certain conjectured
geometric objects, called motives, lead to Hasse-Weil L-series
that arise from automorphic representations \cite{l79}. At present
very little is known in this direction as far as general structure
theorems are concerned.  There exists, however, a subclass of
varieties for which useful results have been known for some time,
and which turns out to be useful in the present context. In
dimension one this is the class of elliptic curves with complex
multiplication (CM), i.e. such curves admit a symmetry algebra
that is exceptionally large. It was first shown by Deuring in the
fifties \cite{d50s}, following a suggestion of Weil in \cite{w52},
that for such tori with CM the cohomological L-function becomes a
number theoretic object. More precisely, he showed that associated
to the complex multiplication field of the elliptic curve are
algebraic Hecke characters which describe the Hasse-Weil
L-function, much like Weil's Jacobi-sum Gr\"o\ss encharaktere do
in the case of the Fermat varieties. This provides an explicit
description of the L-function for toroidal compactifications.

Complex multiplication is a group property, and it is not obvious
what the most convenient physical generalization of this notion is
for higher dimensional Calabi-Yau varieties. One interesting
attempt in this direction was recently made by Gukov and Vafa
\cite{gv02}, who conjectured that exactly solvable Calabi-Yau
varieties can be characterized in terms of a property of the
intermediate Jacobian described in \cite{m69, pss73, b92} (see
also \cite{y03}). In the present paper we follow a different
approach, which is motivated in part by the results of \cite{su02}
and \cite{lps03}. In \cite{lps03} our focus was on properties of
black hole attractor Calabi-Yau varieties with finite fundamental
group. In an interesting paper Moore \cite{m98} has shown that
attractor varieties with elliptic factors are distinguished by the
fact that they admit complex multiplication. The aim of
\cite{lps03} was to introduce a framework in which the notion of
complex multiplication can be generalized to non-toroidal
Calabi-Yau varieties of arbitrary dimension via abelian varieties
that can be derived from the Calabi-Yau cohomology. Abelian
varieties are natural higher dimensional generalizations of
elliptic curves, and certain types admit complex multiplication.
The link between Calabi-Yau manifolds and abelian varieties
therefore allows us to generalize the elliptic analysis to the
higher dimensional abelian case.

In the most general context, the relation between exactly solvable
Calabi-Yau varieties and complex multiplication very likely goes
beyond abelian varieties and involves the theory of motives with
(potential) complex multiplication. The program of constructing a
satisfactory framework of motives is incomplete at this point,
despite much effort. In this paper we therefore focus on the
simpler case of exactly solvable Calabi-Yau varieties that lead to
motives derived from abelian varieties which admit complex
multiplication. We provide a conceptual explanation of the results
of \cite{rs01} and thereby establish a framework that extends the
analysis of \cite{rs01} to general Calabi-Yau manifolds.

The paper is organized as follows. In Sections 2 and 3 we very
briefly recall the arithmetic and number theoretic concepts that
will be used in the following parts. In Section 4 we discuss two
explicit examples of Fermat type which illustrate the transition
from geometry to number theory in a concrete way. In Section 5 we
discuss Deuring's result, which provides a number theoretic
interpretation of the L-function of arbitrary elliptic curves with
complex multiplication. We use an id\`elic formulation of
Deuring's ideal theoretic analysis because this allows us to
clearly identify the geometric structure that provides the
conceptual basis of this result $-$ the discrete set of torsion
point on the elliptic curves.In Section 6 we review the structure
of higher dimensional abelian varieties with complex
multiplication and show how their L-functions can be expressed in
terms of algebraic Hecke characters. In Section 7 we describe how
one can associate abelian varieties to Calabi-Yau manifolds by
tracing the cohomology of Calabi-Yau varieties to the Jacobians of
curves \cite{lps03}. Section 8 contains an example, and in Section
9 we collect some results from ad\`elic number theory for
convenience.

\vskip .2truein

\section{Arithmetic L-functions}

\subsection{The Hasse-Weil L-function}

The starting point of the arithmetic analysis is the set of Weil
conjectures \cite{w49}, the proof of which was completed by
 Deligne \cite{d74}.
For algebraic varieties the Weil--Deligne result states a number
of structural properties for the congruent zeta function at a
prime number $p$ defined as \beq Z(\XFp, t) \equiv
exp\left(\sum_{r\in \IN} \# \left(\XFpr\right)
\frac{t^r}{r}\right). \lleq{artinzeta} The motivation to arrange
the numbers $N_{p^r}= \# \left(\XFpr\right)$ in this particular
way, rather than a more naive generating function like $\sum_r
N_{r,p}t^r$,  originates from the fact that they often show a
simple behavior, as a result of which the zeta function can be
shown to be a rational function. This was first shown by Artin in
the 1920s for hyperelliptic function fields \cite{ea24} and by
Schmidt for curves of arbitrary genus \cite{fks31, hh34}. Further
experience by Hasse, Weil, and others led to the conjecture that
this phenomenon is more general, culminating in the Weil
conjectures, and Deligne's proof in the 1970s.

The part of the conjectures that is most important for the present
context is that the rational factors of $Z(X/\IF_p,t)$ \beq
Z(\XFp,t)=\frac{\prod_{j=1}^d \cP^{(p)}_{2j-1}(t)}{\prod_{j=0}^d
\cP^{(p)}_{2j}(t)}, \eeq can be written as \beq
\cP_0^{(p)}(t)=1-t,~~~ \cP_{2d}^{(p)}(t)=1-p^dt \eeq and for
$1\leq i \leq 2d-1$ \beq \cP_i^{(p)}(t) = \prod_{j=1}^{b_i}
\left(1-\beta^{(i)}_j(p) t\right), \eeq with algebraic integers
$\b^{(i)}_j(p)$. The degree of the polynomials $\cP_i^{(p)}(t)$ is
given by the Betti numbers of the variety, $b_i={\rm dim~
H}^i_{\rm DeRham}(X)$. The rationality of the zeta function was
first shown by Dwork \cite{bd60} by ad\'elic methods. More details
of the Weil conjectures were briefly described in \cite{rs01}.

 We see from the rationality of the zeta function that the basic information
 of this quantity is parametrized by the cohomology of the variety.
 More precisely, one can show that  the $i$'th polynomial $\cP_i^{(p)}(t)$ is
 associated to the action induced by the Frobenius morphism on the
 $i^{\rm th}$ cohomology group $\rmH^i(X)$. In order to gain insight
 into the arithmetic information encoded in these Frobenius
 actions it is useful to decompose the zeta function of the variety
 into pieces determined by its cohomology. This leads to the
 concept of a local L-function that is associated to the
 polynomials $\cP_i^{(p)}(t)$ via the following definition.

 Let $\cP_i^{(p)}(t)$ be the polynomials
 determined by the rational congruent zeta function over the field
 $\IF_p$. The $i^{\rm th}$ L-function of the variety
 $X$ over $\IF_p$ then is defined via \beq L^{(i)}(X/\IF_p, s) =
 \frac{1}{\cP_i^{(p)}(p^{-s})}.\eeq

 Such L-functions are of interest for a number reasons. One of
these is that often they can be modified by simple factors so that
after analytic continuation they (are conjectured to) satisfy some
type of functional equation.

\subsection{Arithmetic via Jacobi sums}

In the case of weighted projected Brieskorn-Pham varieties it is
possible to provide more insight into the structure of the
L-function polynomials $\cP_d^{(p)}(t)$. In the case of Fermat
hypersurfaces it is an old result by Weil according to which the
cardinalities of the variety in terms of Jacobi sums of finite
fields.

{\bf Theorem.}\cite{w49} {\it Define the number $d=(n,q-1)$ and
the set } \beq \cA_s^{q,n} = \left\{(\a_0,...,\a_s) \in
\mathQ^{s+1}~|~ 0<\a_i<1, d\a_i =0~(\rmmod~1), \sum_i \a_i =0
~(\rmmod~1)\right\}.\eeq {\it Then the number of solutions of the
projective variety} \beq X_{s-1} = \left\{(z_0:z_1:\cdots :z_s)
\in \IP_s~|~\sum_{i=0}^s b_iz_i^n=0\right\} \subset \IP_s \eeq
{\it over the finite field $\IFq$ is given by} \beq N_q(X_{s-1}) =
1 + q+ q^2 + \cdots + q^{s-1} + \sum_{\a \in \cA_s^{q,n}} j(\a)
\prod \bchi_{\a_i}(b_i), \eeq {\it where $d=(n,q-1)$ and } \beq
j_q(\a) = \frac{1}{q-1} \sum_{\stackrel{u_i \in \IF_q}{ u_0 +
\cdots + u_s=0}} \chi_{\a_0}(u_0)\cdots \chi_{\a_s}(u_s), \eeq
{\it with} \beq \chi_{\a_i}(u_i) = e^{2\pi i \a_i m_i}, \eeq {\it
where $m_i$ is determined via $u_i = g^{m_i}$ for any generator $g
\in \IF_q$.}

With these Jacobi sums $j_q(\a_0,...,\a_s)$ one defines the
polynomials \beq \cP_{s-1}^{(q)}(t) = \prod_{\a \in \cA_s^{n,q}}
\left(1-(-1)^{s-1}j_q(\a_0,...,\a_s)\prod_i\bchi_{\a_i}(b_i) t
\right) \eeq and the associated L-function \beq L^{(j)}(X,s) =
\prod_p \frac{1}{\cP_j^{(p)}(p^{-s})} .\eeq

A slight modification of this result is useful even in the case of
weighted projective Brieskorn-Pham varieties because it can be
used to compute the factor of the zeta function coming from the
invariant part of the cohomology, when viewing these spaces as
quotient varieties of projective spaces \cite{y01}.

\vskip .2truein

\section{L-Functions of algebraic number fields}

The surprising aspect of the Hasse-Weil L-function is that it is
determined by another, a priori completely different kind of
L-function that is derived not from a variety but from a number
field. It is this possibility to interpret the cohomological
Hasse-Weil L-function as a field theoretic L-function which
establishes the connection that allows us to derive number fields
$K$ from algebraic varieties $X$.

In the present context the type of L-function that is important is
that of a Hecke L-function determined by a Hecke character, more
precisely an algebraic Hecke character. Following Weil we will see
that the relevant field for Fermat type varieties is the
cyclotomic field extension $\mathQ(\mu_m)$ of the rational field
$\mathQ$ by roots of unity, generated by $\xi=e^{2\pi i/m}$ for
some rational integer $m$. It turns out that these fields fit in
very nicely with the conformal field theory point of view. In
order to see how this works we first describe the concept of Hecke
characters and then explain how the L-function fits into this
framework.

There are many different different definitions of algebraic Hecke
characters, depending on the precise number theoretic framework.
Originally this concept was introduced by Hecke \cite{h18} as
Gr\"o\ss encharaktere of an arbitrary algebraic number field. In
the following Deligne's adaptation of Weil's Gr\"o\ss encharaktere
of type $A_0$ is used \cite{d77}.

{\bf Definition.} {\it Let $\cO_K \subset K$ be the ring of
integers of the number field $K$, $\ffrak \subset \cO_K$ an
integral ideal, and $F$ a field of characteristic zero. Denote by
$\cI_{\ffrak}(K)$ the set of fractional ideals of $K$ that are
prime to $\ffrak$ and denote by $\cI_{\ffrak}^p(K)$ the principal
ideals $(\a)$ of $K$ for which $\a\equiv 1(\rmmod~\ffrak)$. An
algebraic Hecke character modulo $\ffrak$ is a multiplicative
function $\chi$ defined on the ideals $ \cI_{\ffrak}(K)$ for which
the following condition holds. There exists an element in the
integral group ring $\sum n_{\si}\si \in \ZZ[\rmHom(K,\bF)]$,
where $\bF$ is the algebraic closure of $F$, such that if $(\a)
\in \cI_{\ffrak}^p(K)$ then} \beq \chi((\a)) = \prod_{\si}
\si(\a)^{n_{\si}}. \eeq {\it Furthermore there is an integer $w>0$
such that $n_{\si}+n_{\bsi}=w$ for all $\si \in \rmHom(K,\bF)$.
This integer $w$ is called the weight of the character $\chi$.}

Given any such character $\chi$ defined on the ideals of the
algebraic number field $K$ we can follow Hecke and consider a
generalization of the Dirichlet series via the L-function \beq
L(\chi,s)=\prod_{\stackrel{\pfrak \subset \cO_K}{\pfrak ~{\rm
prime}}} \frac{1}{1-\frac{\chi(\pfrak)}{\rmN\pfrak^s}} ~=~
\sum_{\afrak \subset \cO_K}
\frac{\chi(\afrak)}{\rmN\afrak^s},\lleq{heckeL} where the sum runs
through all the ideals. Here $\rmN\pfrak$ denotes the norm of the
ideal $\pfrak$, which is defined as the number of elements in
$\cO_K/\pfrak$. The norm is a multiplicative function, hence it
can be extended to all ideals via the prime ideal decomposition of
a general ideal. If we can deduce from the Hasse-Weil L-function
the particular Hecke character(s) involved we will be able to
derive directly from the variety in an intrinsic way distinguished
number field(s) $K$.

Insight into the nature of number fields can be gained by
recognizing that for certain extensions $K$ of the rational number
$\mathQ$ the higher Legendre symbols provide the characters that
enter the discussion above. Inspection then suggests that we
consider the power residue symbols of cyclotomic fields
$K=\mathQ(\mu_m)$ with integer ring $\cO_K=\ZZmum$. The transition
from the cyclotomic field to the finite fields is provided by the
character which is determined for any algebraic integer $x\in
\ZZmum$ prime to $m$ by the map \beq \chi_{\bullet} (x):
\Ifrak_m(\cO_K) \lra \mathC^{\times},\eeq which is defined on
ideals $\pfrak$ prime to $m$ by sending the prime ideal to the
$m$'th root of unity for which \beq
\pfrak~~\mapsto~~\chi_{\pfrak}(x)=x^{\frac{\rmN\pfrak-1}{m}}
(\rmmod~\pfrak).\eeq Using these characters one can define
Jacobi-sums of rank $r$ for any fixed element $a=(a_1,...,a_r)$ by
setting \beq J_a^{(r)}(\pfrak)=(-1)^{r+1} \sum_{\stackrel{u_i\in
\cO_K/\pfrak}{\sum_i u_i=-1 (\rmmod~\pfrak)}}
\chi_{\pfrak}(u_1)^{a_1}\cdots \chi_{\pfrak}(u_r)^{a_r} \eeq for
prime $\pfrak$.  For non-prime ideals $\afrak \subset \cO_K$ the
sum is generalized via prime decomposition $\afrak = \prod_i
\pfrak_i$ and multiplicativity $J_a(\afrak)=\prod_i
J_a(\pfrak_i)$. Hence we can interpret these Jacobi sums as a map
$J^{(r)}$ of rank $r$ \beq J^{(r)}: \Ifrak_m(\ZZmum) \times
(\ZZ/m\ZZ)^r \lra \mathC^{\times}, \eeq where $\Ifrak_m$ denotes
the ideals prime to $m$. For fixed $\pfrak$ such Jacobi sums
define characters on the group $(\ZZ/m\ZZ)^r$. It can be shown
that for fixed $a\in (\ZZ/m\ZZ)^r$ the Jacobi sum $J_a^{(r)}$
evaluated at principal ideals $(x)$ for $x\equiv 1(\rmmod~m^r)$ is
of the form $x^{S(a)}$, where \beq S(a) =
\sum_{\stackrel{(\ell,m)=1}{\ell~\rmmod~m}}\left[\sum_{i=1}^r
\left< \frac{\ell a_i}{m}\right>\right]\si_{\ell}^{-1},\eeq where
$<x>$ denotes the fractional part of $x$ and $[x]$ describes the
integer part of $x$.

\vskip .2truein

\section{Examples}

\subsection{The elliptic Fermat curve}

In \cite{su02} the elliptic curve defined by the plane cubic torus
\beq C_3 = \{(z_0:z_1:z_2) \in \IP_2~|~ z_0^3 + z_1^3 + z_2^3 =0\}
\lleq{planecubic} was analyzed in some detail.

The zeta function (\ref{artinzeta}) simplifies for curves into the
form \beq Z(X,s) = \prod_{\ZZ \ni p~{\rm good~prime}}
\frac{\cP^{(p)}(p^{-s})}{(1-p^{-s})(1-p^{1-s})} =
\frac{\zeta(s)\zeta(s-1)}{L_{\rmHW}(X,s)}, \eeq written in terms
of the Hasse-Weil L-function defined as \beq L_{\rmHW}(X,s) =
\prod_{\ZZ\ni p~{\rm good~prime}}
\frac{1}{\cP^{(p)}(p^{-s})},\lleq{lhasseweil} and the Riemann zeta
function $\zeta(s) = \prod_p (1-p^{-s})^{-1}$ of the rational
field $\mathQ$.

The Hasse-Weil L-function can be determined via (\ref{artinzeta})
by direct counting of the number of solutions of $C_3/\IF_{p^r}$
over finite extensions $[\IF_{p^r}:\IF_p]$ of the finite fields
$\IF_p$ of prime order $p$. This results in \beq L_{\rmHW}(C_3,s)
= 1 - \frac{2}{4^s} - \frac{1}{7^s} + \frac{5}{13^s} +
\frac{4}{16^s} - \frac{7}{19^s} + \cdots, \eeq leading to the
Hasse-Weil $q-$expansion \beq f_{\rmHW}(C_3,q) = q - 2q^4 - q^7 +
5q^{13} + 4q^{16} - 7q^{19} + \cdots \lleq{hwmodform} It turns out
that this is a modular form of weight 2 and modular level 27 which
can be written as a product of the theta function $\Theta(\tau)$
associated to the string function $c(\tau)$ of the affine SU(2)
Kac-Moody algebra at conformal level $k=1$. More precisely, the
following result was obtained.

{\bf Theorem.}(\cite{su02}) {\it The Mellin transform of the
Hasse-Weil L-function $L_{\rmHW}(C_3,s)$ of the cubic elliptic
curve $C_3 \subset \IP_2$ is a modular form $f_{\rmHW}(C_3,q) \in
S_2(\Gamma_0(27))$ which factors into the product} \beq
f_{\rmHW}(C_3,q) = \Theta(q^3)\Theta(q^9).\eeq {\it Here
$\Theta(\tau)=\eta^3(\tau)c(\tau)$ is the Hecke modular form
associated to the quadratic extension ${\mathbb Q}(\sqrt{3})$ of
the rational field $\mathbb{Q}$, determined by the unique string
function $c(\tau)$ of the affine Kac-Moody SU(2)-algebra at
conformal level $k=1$.}

This establishes that it is possible to derive the modularity of
the underlying string theoretic conformal field theory from the
geometric target space and that the Hasse-Weil L-function admits a
conformal field theoretic interpretation.

The number theoretic interpretation of the Hasse-Weil L-function
is best seen from the expression for the polynomials
$\cP^{(p)}(t)$, which completely determine the congruent zeta
function and the Hasse-Weil L-function of these plane curves, in
terms of the finite field Jacobi sums. For curves this reduces to
\beq \cP^{(p)}(t) = \prod_{\a \in \cA_2^p} (1-j_p(\a)t) \eeq with
\beq j_q(\a) = \frac{1}{q-1} \sum_{\stackrel{u_i \in \IF_q}{ u_0 +
u_1 + u_2=0}} \chi_{\a_0}(u_0) \chi_{\a_1}(u_1)\chi_{\a_2}(u_2).
\eeq

Collecting values at primes is in part easier than direct counting
because the cardinalities of the sets $\cA_2^p$ are easy to
control. For the first few primes the results are collected in the
following Table 1, where the zeroes follow immediately from the
fact that the corresponding sets $\cA_2^p$ are empty.
\begin{center}
\begin{tabular}{l| c c c c c c c}
$q$ &2  &3   &5  &7             &9  &11  &13\tabroom \\

\hline $j_q\left({\tiny \frac{1}{3}, \frac{1}{3}, \frac{1}{3}}
\right)$
    &0  &0   &0  &$2+3\xi_3^2$  &   &0   &$-1+3\xi_3^2$\tabroom \\
$j_q\left({\tiny \frac{2}{3}, \frac{2}{3}, \frac{2}{3}} \right)$
    &0  &0   &0  &$2+3\xi_3$    &   &0   &$-1+3\xi_3$\tabroom \\
\hline
\end{tabular}
\end{center}

\vskip .1truein {\bf Table 1.}~{\it Finite field Jacobi sums of
the elliptic cubic curve $C_3$ at the lower rational primes.}

By translating the finite field Jacobi-sums into Jacobi-sum type
Hecke characters $J_{(i,i,i)}(\afrak)$ of the cyclotomic field
${\mathbb Q}(\mu_3)$ one can write the geometric Hasse-Weil
L-function as a number theoretic object associated to this field.
Applied to the field $\mathQ(\mu_3)$ this procedure leads to the
number theoretic representation of the Hasse-Weil L-function of
the plane cubic curve as
 \beq L_{\rmHW}(E,s) =  L_H(J_{(1,1,1)},s)L_H(J_{(2,2,2)},s).\eeq

\subsection{The quintic threefold}

Consider the Calabi-Yau variety defined by the Fermat quintic
hypersurface in ordinary projective fourspace $\IP_4$ defined by
\beq X = \left\{(x_0:\cdots :x_4) \in \IP_4~{\Big |}~\sum_{i=0}^4
x_i^5=0\right\}.\eeq It follows from Lefshetz's hyperplane theorem
that the cohomology below the middle dimension is inherited from
the ambient space. Thus we have $h^{1,0}=0=h^{0,1}$ and
$h^{1,1}=1$ while $h^{2,1}=101$ follows from counting monomials of
degree five. For the smooth Fermat quintic the zeta function
simplifies to the expression \beq Z(X/\IF_p,t) =
\frac{\cP_3^{(p)}(t)}{(1-t)(1-pt)(1-p^2t)(1-p^3t)},\eeq
 where the numerator is given by the polynomial
$\cP_3^{(p)}(t)=\prod_{i=1}^{204} (1-\beta_i^{(3)}(p)t)$ which
takes the form  \beq \cP_3^{(p)}(t)= \prod_{\a \in \cA}
\left(1-j_p(\a) t \right).\eeq This expression involves the
following ingredients. Define $\delta = (5,p-1)$ and rational
numbers $\a_i$ via $\delta \a_i\equiv 0(\rmmod~1)$. The set $\cA$
is defined as \beq \cA=\{\a=(\a_0,...,\a_4) ~|~0<\a_i <1,~\delta
\a_i\equiv 0(\rmmod~1), \sum_i \a_i=0(\rmmod~1)\}.\eeq Defining
the characters $\chi_{\a_i}\in \hat{\IF_p}$ in the dual of $\IF_p$
as $\chi_{\a_i}(u_i)=exp(2\pi i\a_i s_i)$ with $u_i=g^{s_i}$ for a
generating element $g \in \IF_p$, the factor $j_p(\a)$ finally is
determined as \beq j_p(\a) =\frac{1}{p-1}\sum_{\sum_i u_i=0}
\prod_{i=0}^4 \chi_{\a_i}(u_i).\eeq

We thus see that the congruent zeta function leads to the
Hasse-Weil L-function associated to a Calabi-Yau threefold \beq
L_{\rm HW}(X,s) = \prod_{p \in P(X)} \prod_{\a \in \cA}
\left(1-\frac{j_p(\a)}{p^s}\right)^{-1},\eeq ignoring the bad
primes, which are irrelevant for our purposes.

\vskip .2truein

\section{L-function of elliptic curves with complex
multiplication}

\subsection{Deuring's decomposition}

Calabi-Yau varieties of Brieskorn-Pham type provide important
examples of exactly solvable string vacua, but not all exact
string models which admit a geometric interpretation lead to such
spaces. The question therefore arises whether there are other
varieties for which a number theoretic decomposition is possible.
This issue was addressed first by Deuring for elliptic curves with
complex multiplication, i.e. elliptic curves $E$ whose
endomorphism algebra $\rmEnd(E)$ is not restricted to
multiplication by integers. He showed that a similar structure of
the Hasse-Weil L-function arises with characters that are
algebraic Hecke characters $\chi$ (more details on the
construction of these characters will be described in the next
subsection). Given any such character defined on the ideals of the
algebraic number field $K$ we can consider Hecke's L-function
(\ref{heckeL}). The result of Deuring then determines the
Hasse-Weil L-function of any elliptic curve with complex
multiplication as a number theoretic object. Even though Deuring's
result is general for CM elliptic curve, the arguments that
establish this identity depend on the behavior of the complex
multiplication field $F$ relative to the field of definition $K$.
The following result describes the simpler situation when the
field $F$ is contained in the field $K$.

{\bf Theorem.} (\cite{d50s}). {\it Let $E/K$ be an elliptic curve
with complex multiplication by the ring of integers $\cO_F$ of the
algebraic field $F$, $\rmEnd(E) \cong \cO_F$. Assume that $F$ is
contained in $K$ and let $\chi$ be the algebraic Hecke character
associated to $E$. Then} \beq L(E/K,s) = L(\chi,s)L({\bar
\chi},s).\lleq{ell-hasse-weil-hecke}

This result shows that for any elliptic curve with complex
multiplication the geometric Hasse-Weil L-function leads to
L-functions associated to algebraic number fields. In the case of
the Fermat curve this number field is the fusion field, determined
by the quantum dimensions of the scaling fields of the underlying
conformal field theory. Deuring's result shows that this can be
generalized to any elliptic curve with complex multiplication.

This then leads to the idea that perhaps quite generally for
elliptic curves with CM the fusion field of the rational conformal
field theory can be obtained via the Hecke L-function
interpretation of the Hasse-Weil L-function. We therefore should
ask what precisely the underlying structure is that is responsible
for the existence of these characters.

\vskip .2truein

\subsection{Character construction from elliptic curves}

Our goal in this section is to describe the conceptual origin of
the Hecke characters that appear in Deuring's result, and which
turn out to provide one of the links from geometry to conformal
field theory. Even though its explication is not absolutely
necessary for the logic of our argument, it is worthwhile to
briefly outline the key elements that facilitate the transition
from the elliptic curve to the algebraic Hecke characters, because
it illuminates the number theoretic nature of elliptic curves, and
more generally, abelian varieties, with complex multiplication. It
is this number theoretic structure that lies at the heart of the
results obtained so far concerning a direct relation between
exactly solvable models and geometric string vacua.

The construction is based on the action of the Artin symbols on
the points of finite order of the abelian variety, and therefore
combines both geometric and number theoretic aspects. While the
geometric concepts are completely parallel in the 1-dimensional
and the higher dimensional case, the transition from elliptic
curves to more general abelian varieties introduces a number of
technical complications that do not help to illuminate the essence
of the argument. It is therefore useful to consider first elliptic
curves. More details can be found in \cite{js97, js94, gs71a}.

The starting point of the analysis of the action of the Artin
symbol on the torsion points is the main theorem of complex
multiplication. Before stating it we review some concepts. Let
$E/K$ be an elliptic curve over a number field $K$ and let \bea f:
\mathC/\Lambda &\lra & E \otimes \mathC \nn \\ z &\mapsto &
({\wp}(z,\Lambda), {\wp}'(z,\Lambda)) \eea describe the analytic
representation of $E\otimes \mathC$ for a lattice $\Lambda \subset
\mathC$ via the Weierstrass $\wp-$function. Such a representation
always exists according to the uniformization theorem for elliptic
curves, and hence we obtain a polynomial description, given by the
Weierstrass equation \beq E:~~y^2 = 4x^3 - g_2(\Lambda)x -
g_3(\Lambda).\lleq{weierstrasseq} Assume that $E$ has complex
multiplication by the field $F$, i.e. there exists an isomorphism
\beq \theta: F \lra \rmEnd(E) \otimes \mathQ. \eeq It follows that
$F$ must be an imaginary quadratic field (see e.g. \cite{s66}). If
an elliptic curve has CM by a field $F$ then $E$ can be
constructed as $\mathC/\afrak$, where $\afrak$ is a
$\mathZ-$lattice. More precisely, there is a short exact sequence
\beq 0 \lra \afrak \lra \mathC \lra E \lra 0. \eeq

Denote by $E[m]$ the set of finite order on $E$, i.e. the kernel
of the multiplication map by $m$ \beq E[m] = \{z \in E~|~mz =0\},
\eeq and let $E_{\rmtor}$ denote the torsion points of $E$, i.e.
the collection of all points of arbitrary finite order.  The
analytic parametrization then restricts to a map, also denoted by
$f$, \beq f: F/\afrak \lra E_{\rmtor}. \eeq

Artin symbols are particular elements in the Galois group
$\rmGal(F_{\rmab}/F)$ of the maximal abelian extension $F_{\rmab}$
of $F$. They are most efficiently described in terms of id\`eles
$\mathA_F^{\times}$ \cite{c36-40}, which we can view as the
multiplicative subgroup of the ring of ad\`eles $\mathA_F$
\cite{aw45}\fnote{1}{A brief review of ad\`eles, id\`eles, and
class field theory is contained in the appendix.}. Class field
theory says that there exists a homomorphism \beq
\mathA_F^{\times} \lra \rmGal(F_{\rmab}/F) \lleq{reciproc} from
the group of id\`eles onto the Galois group of the maximal abelian
extension. Hence for any finite extension $F$ and for any $\si \in
\rmGal(\mathC/F)$ there exists an id\`ele $x\in \mathA_F^{\times}$
such that \beq \si|_{F_{\rmab}} \in \rmGal(F_{\rmab}/F) \eeq is
determined by $x$. This element is usually denoted by $[x,F]$, and
can be constructed for any finite abelian extension $L/F$ in terms
of the Artin symbol associated to the ideal $(x)$ that can be
associated to the id\`ele by defining \beq (x) = \prod_{\pfrak}
\pfrak^{\rmord_{\pfrak}x_{\pfrak}},\eeq where $\pfrak$ are prime
ideals in $F$ and $x_{\pfrak} \in K_{\pfrak}$ denotes the
components of the id\`ele in the completion of $K$ defined by the
norm $|\cdot|_{\pfrak}$ associated to $\pfrak$. Restricted to the
extension $L$ the map $[x,F]$ then is defined as \beq [x,F]|_L =
\si_{(x)}, \eeq where for any prime $\Pfrak$ of $L$ that divides
the prime ideal $\pfrak$ of $F$ the Artin symbol $\si_{\Pfrak}$ is
defined as \beq \si_{\Pfrak}(x) = x^{\rmN\pfrak} ~\rmmod~\Pfrak.
\lleq{artinsymbol}

The homomorphism (\ref{reciproc}) has a nontrivial kernel given by
the connected component of the id\`ele class group $C_F =
\mathA_F^{\times}/F^{\times}$. Denoting this connected component
by $D_F$ then leads to an identification \beq C_F/D_F
\stackrel{\cong}{\lra} \rmGal(F_{\rmab}/F). \eeq

This result shows that it makes sense to ask what the action is of
an ideal (class) on an elliptic curve. To answer this question we
need a notion of multiplication of a fractional ideal $\afrak$ of
$F$ by an id\`ele $x\in \mathA_F^{\times}$. The idea here is to
use the fractional ideal $(x)$ constructed from the id\`ele $x$
and then define the product via this ideal \beq x\afrak = (x)
\afrak,~~~~x\in \mathA_F^{\times},~\afrak \subset F.\eeq We also
need the inverse $\afrak^{-1}$ of an ideal $\afrak$ in $F$, which
can be defined as \beq \afrak^{-1} = \{x\in F~|~x\afrak \subset
\cO_F\}. \eeq

If we now think of $E$ in terms of its analytic representation
$\mathC/\afrak$ with $\afrak \subset \cO_F$, then for any non-zero
$\bfrak \subset \cO_F$ we can ask what the action of the ideal
(class) does on the elliptic curve, i.e. we can consider $E'=
\mathC/\bfrak^{-1}\afrak$ (here the appearance of the inverse of
$\bfrak$ is conventional). If, on the other hand, we think of $E$
as defined by a polynomial then we can ask how the transformation
of the coefficients in the polynomial affects the curve. This
leads to the notion of the $\si-$transform $E^{\si}$ of an
elliptic curve $E$ for automorphisms of $\mathC$, which is defined
via an action of the map on the coefficients of the curve. In the
case of the Weierstrass representation (\ref{weierstrasseq}) the
$\si-$transform is described by the curve \beq E^{\si}:~~y^2=4x^3
-g_2^{\si}x - g_3^{\si}.\eeq

The main result in the theory of complex multiplication for
elliptic curves answers the question what the relation is between
these two geometric transformations of an elliptic curve, and what
this relation implies for the torsion points.

{\bf Theorem.}~ {\it Let $E$ be an elliptic curve described by an
isomorphism} \beq f: \mathC/\afrak \stackrel{\cong}{\lra}
E(\mathC), \eeq {\it where $\afrak$ is a fractional ideal in $F$.
Assume that $E$ admits complex multiplication by the number field
$F$, $\rmEnd(E) \cong \cO_F$, and let $\si \in \rmGal(\mathC/F) =
\rmAut_F(\mathC)$ and $x\in \mathA_F^{\times}$ be such that} \beq
\si|_{F_{\rmab}} = [x,F]|_{F_{\rmab}}.\eeq {\it Then there exists
an analytic isomorphism} \beq f': \mathC/x^{-1}\afrak
\stackrel{\cong}{\lra} E(\mathC)^{\si} \eeq {\it such that the
following diagram is commutative} \beq \matrix{ F/\afrak
&\stackrel{x^{-1}}{\lra} &F/x^{-1}\afrak \cr
         &                        &         \cr
f\downarrow &                     &\downarrow f' \cr
         &                        &          \cr
E(\mathC) &\stackrel{\si}{\lra}   &E(\mathC)^{\si}. \cr}
\lleq{silv-ellipticCM}

Put slightly different, we can take any element $v\in F/\afrak$,
consider the corresponding torsion point $f(v)$, and ask what the
action of the automorphism $\si$ does. This means that we want to
know what $f(v)^{[x,F]}$ is. The diagram (\ref{silv-ellipticCM})
then says that the image under the Artin symbol is given by
 \beq f(a)^{[x,F]} =
f'(x^{-1}a),~~~~\forall a\in F/\afrak,~~ x\in
\mathA_F^{\times}.\eeq

This result by A. Robert is foundational because it implies the
highlights of the classical theory of complex multiplication, such
as the construction of the Hilbert class field, as well as the
maximal abelian extension of quadratic imaginary fields (see e.g.
\cite{gs71a, js94}). A discussion of these standard results in a
physical context can be found in \cite{m98,lps03}. The main
ingredient of the proof of this theorem is the Kronecker
congruence relation in a form given first by Hasse \cite{h27}. It
states that for an extension $L/F$ the Artin symbol
(\ref{artinsymbol}) of a prime $L\supset \Pfrak|\pfrak$ acts on
the $j-$invariant of the elliptic curve as \beq
\si_{\Pfrak}(j(\afrak)) = j(\pfrak^{-1}\afrak).\eeq This leads to
the rhs of the diagram because the $j-$invariant is an isomorphism
invariant for elliptic curves over algebraically closed fields.

The construction of the algebraic Hecke character is now a
two-step procedure. The first ingredient is a map $\a$ constructed
as follows. For any finite extension $L/F$ an id\`elic norm map
\beq \rmN^L_F: \mathA_L^{\times} \lra \mathA_F^{\times}, \eeq can
be defined by specifying what the $v^{\rm th}$ component is of the
image id\`ele, where $v$ runs through the finite primes as well as
the infinite primes, which are associated to the embeddings of the
number field. For $x\in \mathA_K^{\times}$ one sets \beq
(\rmN^L_Fx)_v = \prod_{w|v}\rmN^{L_w}_{F_v}x_w, \eeq where $L_w$
and $F_v$ are completions of the fields $L$ and $F$ at the primes
$w$ and $v$ respectively (see the appendix for more details on
ad\`eles). Let further $F^{\times}$ denote the invertible elements
of $F$.

{\bf Theorem.} ~{\it Let $E/K$ be an elliptic curve with
parametrization} \beq f: \mathC/\afrak \lra E(\mathC), \eeq {\it
where $\afrak \subset F$ is a fractional ideal. Assume that $E$
has complex multiplication by the ring of integers $\cO_F$ of $F$,
and that $F\subset K$. Then $K(E_{\rmtor})$ is abelian over $K$.
Let further $x\in \mathA_K^{\times}$ be an id\`ele of $K$, and $
y=\rmN_{K/F}x$. Then there exists a unique $\a = \a_{E/K}(x) \in
F^{\times}$ with the following properties: \hfill \break 1) $\a
\cO_F = (y)$, where $(y) \subset F$ is the ideal of $y$. \hfill
\break 2) For any fractional ideal $\afrak \subset F$ and any
analytic isomorphism $f$ the following diagram is commutative}
\beq \matrix{ F/\afrak &\stackrel{\a y^{-1}}{\lra} &F/\afrak \cr
          &                            &         \cr
f\downarrow  &                        &\downarrow f \cr
          &                            &   \cr
E(K_{\rmab}) &\stackrel{[x,K]}{\lra}   &E(K_{\rmab}) \cr } \eeq

This result leads to a map $\a: \mathA_K^{\times} \lra F^{\times}$
which does not yet define an algebraic Hecke character because it
is not trivial on the units. But it can be modified in a way such
that an appropriate character emerges.

{\bf Theorem.}~{\it Let $E/K$ be an elliptic curve with complex
multiplication by the ring of integers $\cO_F$ of $F$ and assume
$F\subset K$. Let}  \beq \a_{E/K}: \mathA_K^{\times} \lra
F^{\times} \eeq {\it be the map described above. For any id\`ele
$x\in \mathA_F^{\times}$, let $x_{\infty} \in \mathC^{\times}$ be
the component of $x$ corresponding to the unique archimedean
absolute value of $F$. Define a map} \bea \psi_{E/K}:
\mathA_K^{\times} &\lra & \mathC^{\times} \nn \\ x &\mapsto
&\a_{E/K}(x) \rmN^K_F(x^{-1})_{\infty}.\eea {\it Then the
following hold: \hfill \break 1) $\psi_{E/K}$ is a Gr\"o\ss
encharakter. \hfill \break 2) Let $\Pfrak$ be a prime of $K$. Then
$\psi_{E/K}$ is unramified at $\Pfrak$ if and only if $E$ has good
reduction at $\Pfrak$. (A Gro\" \ss encharakter $\psi:
\mathA_K^{\times} \lra \mathC^{\times}$ is said to be unramified
at $\Pfrak$ if $\psi(\cO_{\Pfrak}^{\times})=1$.)}

It is the L-function $L(\psi_{E/K},s)$ of this algebraic Hecke
character $\psi_{E/K}$ which provides the number theoretic
description of the Hasse-Weil L-function described above in eq.
(\ref{ell-hasse-weil-hecke}) (it can be shown that the id\`elic
version of the Hecke characters reduces to Deligne's formulation
described in the first part of this section \cite{n90}).

In summary, the important conceptual point here is that the
arithmetic structure of the torsion points of elliptic curves with
complex multiplication carries the essential information which
turns the geometric L-function into a number theoretic object. It
is this number theoretic structure which allows the translation of
geometric properties into conformal field theoretic objects (we
will return to this aspect further below).

In the formulation above we have made the simplifying assumption
that the complex multiplication field $F$ is contained in the
field of definition $K$. A similar analysis goes through when this
assumption is dropped, as has been shown by Deuring in the 1950s
in his important sequence of papers \cite{d50s}, in which he
developed the algebraic approach to complex multiplication used
here (see also \cite{d49}). In particular the first paper in this
series remains an illuminating reference about the relation
between these different L-functions up to finitely many primes.

The generalization to higher dimensional abelian varieties is
possible, but complicated by the fact that the number theoretic
structure arises from the so-called reflex field of the complex
multiplication field $F$, as will become clear further below.

\vskip .2truein

\section{Abelian varieties}

\subsection{General definition}

We first review some relevant concepts in the context of abelian
varieties. An abelian variety over some number field $K$ is a
smooth, geometrically connected, projective variety, which is also
an algebraic group with the group law $A\times A \lra A$  defined
over $K$. A concrete way to construct such manifolds is via
complex tori $\mathC^n/\Lambda$ with respect to some lattice
$\Lambda \subset \mathC^n$, or, put differently, via an exact
sequence \beq 0 \lra \Lambda \lra \mathC^n \stackrel{f}{\lra} A
\lra 0, \eeq where $f$ is a holomorphic map. The lattice $\Lambda$
is not necessarily integral and admits a Riemann form, which is
defined as an $\IR$-bilinear form $<,>$ on $\mathC^n$ such that
the following hold: \hfill \break (1)~~$<x,y>$ takes integral
values for all $x,y\in \Lambda$; \hfill \break
(2)~~$<x,y>=-<y,x>$; \hfill \break (3)~~$<x,iy>$ is a symmetric
and positive definite form in $x,y$. \hfill \break The result then
is that a complex torus $\mathC^n/\Lambda$ has the structure of an
abelian variety if and only if there exists a non-degenerate
Riemann form on $\mathC^n/\Lambda$.

\subsection{Abelian varieties of CM type}

A special class of abelian varieties are those of complex
multiplication (CM) type. These are varieties which admit
automorphism groups that are larger than those of general abelian
manifolds. The reason why CM type varieties are special is because
certain number theoretic questions can be addressed in a
systematic fashion for this class. The first to discover this was
Weil \cite{w52} in the context of Fermat type hypersurfaces. The
fact that this relation can be traced to the property of CM for
abelian varieties was first shown by Deuring in the context of
elliptic curves, following a suggestion by Weil. This was later
generalized conditionally to higher dimensions by Taniyama and
Shimura \cite{t57, st61}, Serre and Tate \cite{st68}, and Shimura
\cite{gs71a, gs71b}.

Consider a number field $F$ over the rational numbers $\mathQ$ and
denote by $[F:\mathQ]$ the degree of the field $F$ over $\mathQ$,
i.e. the dimension of $F$ over the subfield $\mathQ$. An abelian
variety $A$ of dimension $n$ is called a CM$-$variety if there
exists an algebraic number field $F$ of degree $[F:\mathQ]=2n$
over the rational numbers $\mathQ$ which can be embedded into the
endomorphism algebra $\rmEnd(A) \otimes \mathQ$ of the variety.
More precisely, a CM-variety is a triplet $(A,\theta, F)$, where
\beq \theta: F \lra \rmEnd(A) \otimes \mathQ \eeq describes the
embedding of $F$. It follows from this that the field $F$
necessarily is a CM field, i.e. a totally imaginary quadratic
extension of a totally real field. The important ingredient here
is that the restriction to $\theta(F) \subset \rmEnd(A)\otimes
\mathQ$ is equivalent to the direct sum of $n$ isomorphisms
$\vphi_1,...,\vphi_n \in \rmIso(F,\mathC)$ such that
$\rmIso(F,\mathC) = \{\vphi_1,...,\vphi_n, \rho\vphi_1,....,\rho
\vphi_n\}$, where $\rho$ denotes complex conjugation. These
considerations suggest calling the pair $(F,\{\vphi_i\})$ a
CM-type. In principle we can think of the CM type as an abstract
representation defined by some matrix \beq \Phi(a) =
\left(\matrix{a^{\vphi_1} &        &  & \cr
                                    &\ddots  &  & \cr
                                    &        &a^{\vphi_n}
                                    \cr}\right),~~~{\rm for}~a\in F,
\lleq{cm-structure} but in the present context $(F,\Phi =
\{\vphi_i\})$ describes the CM-type of a CM-variety
$(A,\theta,F)$.

It is possible to prescribe the CM structure and construct an
abelian variety with that given structure by constructing a
diagram of the following form \beq \matrix{ 0 &\lra  &\afrak &\lra
&F_{\mathR} &\lra
                           &F_{\mathR}/\afrak  &\lra  &0  \cr
           &               &                   &      &
                           &                   &      &   \cr
           &      &\downarrow   &   &\downarrow~u  &
                           &\downarrow         &      &  \cr
           &               &                   &      &
                           &                   &      & \cr
        0 &\lra  &\Lambda      &\lra   &\mathC^n
        &\stackrel{f}{\lra}
                           &A                  &\lra  &0, \cr}
\lleq{constructA} where $u$ is the map  \bea u: F_{\mathR} &\lra &
\mathC^n \nn
\\ a &\mapsto & \left(\matrix{a^{\vphi_1}\cr \vdots \cr
a^{\vphi_n}\cr} \right),\eea defined as an $\mathR-$linear
extension on $F$, and $\afrak$ is the preimage of $u$ of the
lattice $\Lambda$. The abelian variety is thereby obtained as the
quotient $F_{\mathR}/\afrak$ of $F_{\mathR} = F\otimes_{\rho}
\mathR$, with $\rho$ denoting complex conjugation, by an ideal in
$F$, with a complex structure determined by $u$, and an embedding
$\theta: F\lra \rmEnd(A)\otimes \mathQ$ given by
(\ref{cm-structure}).

Concrete examples of these concepts which have been discussed in
\cite{rs01} in the context of the Calabi-Yau/conformal field
theory relation are varieties which have complex multiplication by
a cyclotomic field $F=\mathQ(\mu_n)$, where $\mu_n$ denotes the
cyclic group generated by a nontrivial $n$'th root of unity
$\xi_n$. The degree of $\mathQ(\mu_n)$ is given by
$[\mathQ(\mu_n):\mathQ]=\phi(n)$, where $\phi(n)=\# \{m\in
\IN~|~m<n, ~\rmgcd(m,n)=1\}$ is the Euler function. Hence the
abelian varieties encountered have complex dimension $\rmdim~A =
\phi(n)/2$.

The simplest examples of abelian CM varieties are elliptic curves
with complex multiplication, as discussed above.  More
interestingly for the present context, they occur in the context
of higher dimensional Calabi-Yau varieties via the Shioda-Katsura
decomposition of the cohomology. Further below we briefly review
the reduction of the cohomology of the Brieskorn-Pham varieties to
that generated by curves and then analyze the structure of the
resulting weighted curve Jacobians.

\subsection{L-Function of abelian varieties with complex
multiplication}

In this section we describe the generalization of the conceptual
framework underlying the number theoretic interpretation of the
zeta function of abelian varieties with complex multiplication.
Our goal is to detail the underlying structure that explains this
phenomenon. As in the case of elliptic curves the main objects
that provide the transition from the discrete geometry of the
variety to number theory are the torsion points on the abelian
variety, i.e. the points in the kernel of a multiplication map $n:
A \lra A$, analogous to the corresponding map on the elliptic
curves.

The general concept of a geometric L-function is derived from the
reduction of a variety over discrete fields $\IF_q$ of order $q$.
One way to think about this structure is by considering the fields
$\IF_q$ as residue fields $\cO_K/\pfrak$, generated by prime
ideals $\pfrak$ in the ring of integers $\cO_K$ of some algebraic
number field $K$. Denoting the residue field produced by $\pfrak$
as $K(\pfrak)$ and the reduced variety by $X(\pfrak)$ one can
define the local zeta function as \beq Z(X,\pfrak,s) :=
Z\left(X(\pfrak)/K(\pfrak),t=\rmN \pfrak^{-s}\right).\eeq By
combining these local zeta functions for all prime ideals one
obtains the global zeta function \beq Z(X/K,s) = \prod_{\pfrak
\subset \cO_K} Z(X,\pfrak,s)\eeq of the variety $X$ defined over
the number field $K$.

When the variety has complex multiplication with respect to some
number field $F$ the zeta function admits a number theoretic
interpretation which generalizes the results of Deuring for
elliptic curves with complex multiplication. Associated to the
field $F$ are again Gr\"o\ss encharaktere $\chi_i$, $i=1,...,n$
which lead to Hecke L-functions $L(\chi_i,s)$. The zeta function
of the abelian variety with complex multiplication then is
described by these Hecke L-functions.

{\bf Theorem.} (see \cite{gs71b}) ~{\it Let the abelian CM-variety
$(A,\theta,F)$ be defined over an algebraic number field $K$ of
finite degree. Then the zeta function of $A$ over $K$ coincides
exactly with the product} \beq \prod_{i=1}^n
L(\chi_i,s)L(\bchi_i,s),\eeq {\it where the $\chi_i$ are Gr\"o\ss
encharaktere, and $\bchi_i$ is the complex conjugate of $\chi_i$.}

This result was first shown in a conditional formulation by
Taniyama and Shimura, and Serre and Tate. It shows that if we
could recover abelian varieties from Calabi-Yau manifolds then we
could generalize to higher dimensions the ideas about the CY/CFT
relation formulated in the previous section for elliptic curves.

\subsection{Character construction from abelian varieties}

The character construction from higher dimensional abelian
varieties differs somewhat from that of elliptic curves because of
the emergence of the reflex type, denoted here by
$(\Fhat,\Phihat=\{\vphihat_i\})$ of the complex multiplication
type $(F,\Phi=\{\vphi_i\})$. This reflex field is defined by
adjoining to $F$ all traces determined by the CM type of $F$, i.e.
$\Fhat = F(\{\sum_i x^{\vphi_i}~|~x\in F\})$. To define the reflex
type $\Phihat$ consider a Galois extension $L/\mathQ$ over the
rationals that contains the CM field $F$. Denote by $S$ the subset
of all those elements of the Galois group $\rmGal(L/\mathQ)$ of
$L$ that induce some $\vphi_i$ on $F$ and define further \bea
S^{-1} &=& \left\{\si^{-1}~|~\si \in S\right\} \nn \\ H &=&
\left\{\gamma \in \rmGal(L/\mathQ)~|~\gamma S^{-1} =
S^{-1}\right\}. \eea Then the reflex type of $(F,\Phi)$ is
completed by defining the maps \beq \vphihat_i: \Fhat \lra \mathC
\eeq as those that are obtained from $S^{-1}$. In the one
dimensional case the discussion simplifies because one has $\Fhat
= F$.

Consider an abelian variety $A/K$ of dimension $n$ defined over a
number field $K$ with complex multiplication, i.e. with an
embedding $\theta: F \lra \rmEnd(A) \otimes \mathQ$. Denote this
structure by $(A/K, \theta, F)$ with type
$(F,\Phi=\{\vphi_i\}_{i=1,...,n})$. The appearance of $\Fhat$
leads to a modification of the norm map that appears in the
elliptic construction of the character. To simplify the discussion
we assume that\fnote{2}{This condition generalizes the assumption
in the elliptic case that the CM field is contained in the field
of definition.} $\Fhat \subset K$. We can therefore consider the
id\`elic norm map $\rmN^K_{\Fhat}: ~\mathA_K^{\times} \lra
\mathA_{\Fhat}^{\times}$, defined earlier, and compose it with the
determinant map \beq \delta: \mathA_{\Fhat}^{\times} \lra
\mathA_F^{\times}, \eeq defined as the continuous extension of the
determinant of the reflex type \beq \delta(x) =
\rmdet~\Phihat(x),~~~~\forall x \in \Fhat^{\times}. \eeq

The composition $g:= \delta \circ \rmN^K_{\Fhat}$ of the norm map
and the determinant map provides a map from the $K-$id\`eles to
the $F-$id\`eles \beq g: \mathA_K^{\times} \lra \mathA_F^{\times},
\eeq generalizing the norm map that we encountered in the elliptic
case.

The construction of the character is based again on the same idea
as in the one dimensional case, the action of the id\`elic Artin
symbol $[x,K]$ for $x\in \mathA_K^{\times}$ on the torsion points
and the map $g$ enters this construction in much the same way that
the norm map itself did in the earlier discussion. The main result
of the theory of complex multiplication in the case of abelian
varieties that are relevant to us can be summarized as follows.

{\bf Theorem.}~{\it Let $(F,\Phi)$ be a CM-type, $(\Fhat,\Phihat)$
its reflex, and $\afrak$ a lattice in $F$. Let further
$(A,\theta)$ be of type $(F,\Phi)$, $u: F_{\mathR} \lra \mathC^n$
the map in the diagram (\ref{constructA}), and $f$ the map that
defines $A$ via} \beq 0\lra \Lambda \lra \mathC^n
\stackrel{f}{\lra} A \lra 0.\eeq {\it Further let $\si \in
\rmAut(\mathC/\Fhat)$, $x\in \mathA_{\Fhat}^{\times}$ be an
id\`ele of the reflex field such that} \beq \si|_{\Fhat_{\rmab}} =
[x,\Fhat], \eeq {\it and $g = \delta \circ \rmN^K_{\Fhat}:
\mathA_K^{\times} \lra \mathA_F^{\times}$ the map defined by the
id\`elic extension of the determinant map. Then there is an exact
sequence} \beq 0 \lra u(g(x)^{-1}\afrak) \lra \mathC^n
\stackrel{f'}{\lra} A^{\si} \lra 0 \eeq {\it such that} \beq
f(u(v))^{\si} = f'(u(g(x)^{-1}v))~~~~\forall v \in F/\afrak, \eeq
{\it i.e. there exists a commutative diagram} \beq \matrix{
F/\afrak &\stackrel{g(x)^{-1}}{\lra} &F/g(x)^{-1}\afrak \cr
         &                        &         \cr
\om \downarrow &                  &\downarrow \om' \cr
         &                        &          \cr
A_{\rmtor} &\stackrel{\si}{\lra}   &A_{\rmtor}^{\si} \cr}
\lleq{lang-abelianCM} {\it where $\om = f\circ u$ and
$\om'=f'\circ u$.}

The construction of the algebraic Hecke character associated to
the torsion points of $A$ is now again achieved by constructing an
id\`elic map $\a$ of $K$ in the following way.

{\bf Theorem.} {\it For the map $\om: F_{\mathR} \lra A$ defined
by $\om =f \circ u$ with $F_{\mathR} = F \otimes_{\rho} \mathR$
and $\rho$ denotes complex conjugation, there exists a map} \beq
\a: \mathA_K^{\times} \lra F^{\times} \eeq {\it such that} \beq
\om(v)^{[x,K]} = \om(\a(x) g(x)^{-1}v)~~~\forall x\in
\mathA_K^{\times},~v\in F/\afrak,\eeq {\it which is determined
uniquely by the following properties} \bea \a(x) g(x)^{-1}\afrak
&=& \afrak \nn \\ \a(x)\a(x)^{\rho} &=& \rmN(x), \eea {\it where
$(x)$ is the ideal associated to $x$. Furthermore the kernel of
$\a$ is open in the id\`eles}.

As in the elliptic case we can now define characters $\psi_i$ on
the id\`eles by picking appropriate components from the map that
describes the Artin symbol on the elements of $F/\afrak$. More
precisely, define \beq \psi_i(x) = (\a(x)g(x)^{-1})_{\infty i},
~~~i=1,...,n, \eeq via the infinite primes of the complex
multiplication field $F$.

\vskip .2truein

\section{Abelian varieties from Brieskorn-Pham type hypersurfaces}

In this section we review our construction of abelian varieties
with complex multiplication from Calabi-Yau varieties
\cite{lps03}. The idea is to first reduce the intermediate
cohomology of the Calabi-Yau via the Shioda-Katsura construction
to the cohomology spanned by curves embedded in the manifold and
then to use the results of Faddeev, Gross, Rohrlich, and others to
decompose the Jacobian varieties derived from these curves to find
factors that admit complex multiplication.

\subsection{The Shioda-Katsura decomposition}

The decomposition of the intermediate cohomology of projective
hypersurfaces was first described by Shioda and Katsura
\cite{sk79} and Deligne \cite{d82}. Their analysis can be
generalized to weighted hypersurfaces, in particular the class of
Brieskorn-Pham varieties, perhaps the simplest class of exactly
solvable Calabi-Yau manifolds. This generalization works because
the cohomology $\rmH^3(X)$ for these varieties decomposes into the
monomial part and the part coming from the resolution. The
monomial part of the intermediate cohomology can easily be
obtained from the cohomology of a projective hypersurface of the
same degree by realizing the weighted projective space as a
quotient variety with respect to a product of discrete groups
determined by the weights of the coordinates.

For projective varieties \beq X_d^n =\left\{(z_0,...,z_{n+1})\in
\IP_{n+1}~|~z_0^d + \cdots + z_{n+1}^d = 0\right\} \subset
\IP_{n+1}\eeq the intermediate cohomology can be determined by
lower-dimensional varieties in combination with Tate twists by
reconstructing the higher dimensional variety $X_d^n$ of degree
$d$ and dimension $n$ in terms of lower dimensional varieties
$X_d^r$ and $X_d^s$ of the same degree with $n=r+s$. Briefly, this
works as follows. The decomposition of $X_d^n$ is given as \beq
X_d^{r+s} ~\cong ~ B_{Z_1,Z_2}\left(\left(\pi_Y^{-1}(X_d^r \times
X_d^s)\right)/\mu_d\right), \eeq which involves the following
ingredients.  \hfill \break (1) $\pi_Y^{-1}(X_d^r\times X_d^s)$
denotes the blow-up of $X_d^r\times X_d^s$ along the subvariety
\beq Y = X_d^{r-1}\times X_d^{s-1} \subset X_d^r \times X_d^s.\eeq
The variety $Y$ is determined by the fact that the initial map
which establishes the relation between the three varieties
$X_d^{r+s},X_d^r, X_d^s$ is defined on the ambient spaces as \beq
((x_0,...,x_{r+1}),(y_0,...,y_{s+1}) ~\mapsto~
(x_0y_{s+1},...,x_ry_{s+1},x_{r+1}y_0,...,x_{r+1}y_s).\eeq This
map is not defined on the subvariety $Y$. \hfill \break (2)
$\pi_Y^{-1}(X_d^r \times X_d^s)/\mu_d$ denotes the quotient of the
blow-up $\pi_Y^{-1}(X_d^r \times X_d^s)$ with respect to the
action of
$$\mu_d \ni \xi: ((x_0,...,x_r,x_{r+1}),(y_0,...,y_s,y_{s+1})) ~\mapsto ~
((x_0,...,x_r,\xi x_{r+1}), (y_0,...,y_s,\xi y_{s+1})).$$ (3)
$B_{Z_1,Z_2}\left(\left(\pi_Y^{-1}(X_d^r \times
X_d^s)\right)/\mu_d\right)$ denotes the blow-down in
$\pi_Y^{-1}(X_d^r \times X_d^s)/\mu_d$ of the two subvarieties
$$ Z_1=\IP_r \times X_d^{s-1},~~~~~~Z_2 = X_d^{r-1} \times \IP_s.$$

This construction leads to an iterative decomposition of the
cohomology which takes the following form. Denote the Tate twist
by \beq \rmH^i(X)(j):=\rmH^i(X)\otimes W^{\otimes j}\eeq with
$W=\rmH^2(\IP_1)$ and let $X_d^{r+s}$ be a Fermat variety of
degree $d$ and dimension $r+s$. Then \bea \rmH^{r+s}(X_d^{r+s}) &
\oplus &\sum_{j=1}^r \rmH^{r+s-2j}(X_d^{r-1})(j) \oplus
\sum_{k=1}^s
\rmH^{r+s-2k}(X_d^{s-1})(k) \nn \\
 & & ~~~~ \cong \rmH^{r+s}(X_d^r\times
X_d^s)^{\mu_d} \oplus \rmH^{r+s-2}(X_d^{r-1}\times
X_d^{s-1})(1).\eea This allows us to trace the cohomology of
higher dimensional varieties to that of curves.

Weighted projective hypersurfaces can be viewed as resolved
quotients of hypersurfaces embedded in ordinary projective space.
The resulting cohomology has two components, the invariant part
coming from the projection of the quotient, and the resolution
part. As described in \cite{cls90}, the only singular sets on
arbitrary weighted hypersurface Calabi-Yau threefolds are either
points or curves. The resolution of singular points contributes to
the even cohomology group $\rmH^2(X)$ of the variety, but does not
contribute to the middle-dimensional cohomology group $H^3(X)$.
Hence we need to be concerned only with the resolution of curves
(see e.g. \cite{s87}). This can be described for general CY
hypersurface threefolds as follows. If a discrete symmetry group
$\mathZ/n\mathZ$ of order $n$ acting on the threefold leaves
invariant a curve then the normal bundle has fibres $\mathC_2$ and
the discrete group induces an action on these fibres which can be
described by a matrix \beq \left(\matrix{\a^{mq} &0 \cr 0
&\a^m\cr}\right),\eeq where $\a$ is an $n$'th root of unity and
$(q,n)$ have no common divisor. The quotient
$\mathC_2/(\mathZ/n\mathZ)$ by this action has an isolated
singularity which can be described as the singular set of the
surface in $\mathC_3$ given by the equation \beq
S=\{(z_1,z_2,z_3)\in \mathC_3~|~z_3^n=z_1z_2^{n-q}\}.\eeq The
resolution of such a singularity is completely determined by the
type $(n,q)$ of the action by computing the continued fraction of
$\frac{n}{q}$ \beq \frac{n}{q}= b_1 - \frac{1}{b_2 -
\frac{1}{\ddots - \frac{1}{b_s}}} \equiv [b_1,...,b_s].\eeq The
numbers $b_i$ specify completely the plumbing process that
replaces the singularity and in particular determine the
additional generator to the cohomology $\rmH^*(X)$ because the
number of $\IP_1$s introduced in this process is precisely the
number of steps needed in the evaluation of
$\frac{n}{q}=[b_1,...,b_s]$. This can be traced to the fact that
the singularity is resolved by a bundle which is constructed out
of $s+1$ patches with $s$ transition functions that are specified
by the numbers $b_i$. Each of these gluing steps introduces a
sphere, which in turn supports a (1,1)-form. The intersection
properties of these 2-spheres are described by Hirzebruch-Jung
trees, which for a $\mathZ/n\mathZ$ action is just an $SU(n+1)$
Dynkin diagram, while the numbers $b_i$ describe the intersection
numbers. We see from this that the resolution of a curve of genus
$g$ thus introduces $s$ additional generators to the second
cohomology group $\rmH^2(X)$, and $g\times s$ generators to the
intermediate cohomology $\rmH^3(X)$.

Hence we see that the cohomology of weighted hypersurfaces is
determined completely by the cohomology of curves. Because the
Jacobian variety is the basic geometric invariant of a smooth
projective curve this says that for weighted hypersurfaces the
main cohomological structure is carried by their embedded curves.

\subsection{Cohomology of weighted curves}

For smooth algebraic curves $C$ of genus $g$ the de Rham
cohomology group $\rmH^1_{\rmdR}(C)$ decomposes (over the complex
number field $\mathC$) as \beq \rmH^1_{\rmdR}(C)~\cong~\rmH^0(C,
\Om^1) \oplus \rmH^1(C,\cO).\lleq{hodge-split} The Jacobian $J(C)$
of a curve $C$ of genus $g$ can be identified with \beq
J(C)=\mathC^g/\Lambda,\eeq where $\Lambda$ is the period lattice
\beq \Lambda:= \left\{\left(\dots,\int_a \om_i,\dots
\right)_{i=1,...,g}~{\Big|}~ a \in \rmH_1(C,\mathZ),~ \om_i \in
\rmH^0(C,\Om^1) \right\},\eeq where the $\om_i$ form a basis.
Given a fixed point $p_0\in C$ on the curve there is a canonical
map from the curve to the Jacobian, called the Abel-Jacobi map
\beq \Psi: C \lra J(C), \eeq defined as \beq p \mapsto
\left(\dots, \int_{p_0}^p \om_i,\dots\right) \rmmod~ \Lambda .\eeq

We are interested in curves of Brieskorn-Pham type, i.e. curves of
the form \beq  C_d= \left\{x^d + y^a + z^b =0 \right\} \in
\IP_{(1,k,\ell)}[d],\eeq such that $a=d/k$ and $b=d/\ell$ are
positive rational integers. Without loss of generality we can
assume that $(k,\ell)=1$. The genus of these curves is given by
\beq g(C_d) = \frac{1}{2}(2-\chi) = \frac{(d-k)(d-\ell)+(k\ell
-d)}{2k\ell}. \eeq

For non-degenerate curves in the configurations
$\IP_{(1,k,\ell)}[d]$ the set of forms \beq
\rmH^1_{\rmdR}(\IP_{(1,k,\ell)}[d]) = \left\{ \om_{r,s,t}=
y^{s-1}z^{t-d/\ell}dy~{\Big |}~ r+ks+\ell t = 0~{\rmmod}~d,
~\left(\matrix{1\leq r \leq d-1, \cr ~1\leq s \leq
\frac{d}{k}-1,\cr ~1\leq t \leq \frac{d}{\ell} -1\cr}\right)
\right\}\lleq{weighted-basis} defines a basis for the de Rham
cohomology group $\rmH^1_{\rmdR}(C_d)$ whose Hodge split is given
by \bea \rmH^0\left(C_d,\Om_{\mathC}^1\right) &=&
\left\{\om_{r,s,t}~|~r+ks+\ell t=d\right\} \nn \\
\rmH^1\left(C_d,\cO_{\mathC}\right) &=&
\left\{\om_{r,s,t}~|~r+ks+\ell t=2d\right\}. \eea

In order to show this we view the weighted projective space as the
quotient of projective space with respect to the actions
$\mathZ_k:[0~1~0]$ and $\mathZ_{\ell}: [0~0~1]$, where we use the
abbreviation $\mathZ_k=\mathZ/k\mathZ$ and for any group
$\mathZ_r$ the notation $[a,b,c]$ indicates the action \beq
[a,b,c]:~(x,y,z) \mapsto (\g^ax, \g^by, \g^cz),\eeq where $\g$ is
a generator of the group. This allows us to view the weighted
curve as the quotient of a projective Fermat type curve \beq
\IP_{(1,k,\ell)}[d] = \IP_2[d]/\mathZ_k \times \mathZ_{\ell}:
\left[\matrix{0&1&0\cr 0&0&1\cr}\right].\eeq These weighted curves
are smooth and hence their cohomology is determined by considering
those forms on the projective curve $\IP_2[d]$ which are invariant
with respect to the group actions. A basis for
$\rmH^1_{\rmdR}(\IP_2[d])$ is given by the set of forms \beq
\rmH^1_{\rmdR}(\IP_2[d]) = \left\{ \om_{r,s,t}=
y^{s-1}z^{t-d}dy~{\Large |}~ 0<r,s,t<d,~~r+s+t=0~(\rmmod
~d),~~r,s,t\in \mathN \right\}.\lleq{proj-basis}

Denote the generator of the $\mathZ_k$ action by $\a$ and consider
the induced action on $\om_{r,s,t}$ \beq \mathZ_k:~~\om_{r,s,t}
\mapsto \a^s \om_{r,s,t}.\eeq It follows that the only forms that
descend to the quotient with respect to $\mathZ_k$ are those for
which $s = 0 (\rmmod~k)$. Similarly we denote by $\beta$ the
generator of the action $\mathZ_{\ell}$ and consider the induced
action on the forms $\om_{r,s,t}$ \beq \mathZ_{\ell}:~~\om_{r,s,t}
\mapsto \beta^{t-d} \om_{r,s,t}.\eeq Again we see that the only
forms that descend to the quotient are those for which
$t=0(\rmmod~\ell)$.

\subsection{Abelian varieties from weighted Jacobians}

Jacobian varieties in general are not abelian varieties with
complex multiplication. The question we can ask, however, is
whether the Jacobians of the curves that determine the cohomology
of the Calabi-Yau varieties can be decomposed such that the
individual factors admit complex multiplication by an order of a
number field. In this section we show that this is indeed the case
and therefore we can define the complex multiplication type of a
Calabi-Yau variety in terms of the CM types induced by the
Jacobians of its curves.

It was shown by Faddeev  \cite{f61}\fnote{4}{More accessible are
the references \cite{w76} \cite{g78}, \cite{r78} on the subject.}
that the Jacobian variety $J(C_d)$ of Fermat curves $C_d\subset
\IP_2$ splits into a product of abelian factors $A_{\cO_i}$ \beq
J(C_d) \cong \prod_{\cO_i \in \cI/(\mathZ/d\mathZ)^{\times}}
A_{\cO_i}, \eeq where the set $\cI$ provides a parametrization of
the cohomology of $C_d$, and the sets $\cO_i$ are orbits in $\cI$
of the multiplicative subgroup $(\mathZ/d\mathZ)^{\times}$ of the
group $\mathZ/d\mathZ$. More precisely it was shown that there is
an isogeny \beq i: J(C_d) \lra \prod_{\cO_i \in
\cI/(\mathZ/d\mathZ)^{\times}} A_{\cO_i},\eeq where an isogeny $i:
A \ra B$ between abelian varieties is defined to be a surjective
homomorphism with finite kernel. In the parametrization used in
the previous subsection $\cI$ is the set of triplets $(r,s,t)$ in
(\ref{proj-basis}) and the periods of the Fermat curve have been
computed by Rohrlich \cite{r78} to be \beq \int_{\cA^j\cB^k\kappa}
\om_{r,s,t} = \frac{1}{d} B\left(\frac{s}{d},\frac{t}{d}\right)
(1-\xi^s)(1-\xi^t)\xi^{js+kt}, \eeq where $\xi$ is a primitive
$d-$th root of unity, and \beq B(u,v) = \int_0^1
t^{u-1}(1-v)^{v-1}dt \eeq is the classical beta function.
$\cA,\cB$ are the two automorphism generators
\bea \cA(1,y,z) &=& (1, \xi y, z) \nn \\
\cB(1,y,z) &=& (1, y,\xi z)\eea and $\kappa$ is the generator of
$\rmH_1(C_d)$ as a cyclic module over $\mathZ[\cA,\cB]$. The
period lattice of the Fermat curve therefore is the span of \beq
\left(\dots, \xi^{jr+ks}(1-\xi^r)(1-\xi^s) \frac{1}{d}
B\left(\frac{r}{d},\frac{s}{d}\right), \dots
\right)_{\stackrel{1\leq r,s,t \leq d-1}{r+s+t=d}},~~ \forall
0\leq j,k\leq d-1. \eeq

The abelian factor $A_{[(r,s,t)]}$ associated to the orbit
$\cO_{r,s,t}=[(r,s,t)]$ can be obtained as the quotient \beq
A_{[(r,s,t)]} = \mathC^{\vphi(d_0)/2}/\Lambda_{r,s,t}, \eeq where
$d_0 = d/\rmgcd(r,s,t)$ and the lattice $\Lambda_{r,s,t}$ is
generated by elements of the form \beq
\si_a(z)(1-\xi^{as})(1-\xi^{at})
\frac{1}{d}B\left(\frac{<as>}{d},\frac{<at>}{d}\right), \eeq where
$z\in \mathZ[\mu_{d_0}]$, $\si_a \in
\rmGal(\mathQ(\mu_{d_0})/\mathQ)$ runs through subgroups of the
Galois group of the cyclotomic field $\mathQ(\mu_{d_0})$ and $<x>$
is the smallest integer $0\leq x <1$ congruent to $x$ mod $d$.

Alternatively, the abelian variety $A_d^{r,s,t}$ can be
constructed in a more geometric way as follows. Consider the
orbifold of the Fermat curve $C_d$ with respect to the group
defined as \beq G_d^{r,s,t} = \left\{(\xi_1,\xi_2,\xi_3)\in
\mu_d^3~|~ \xi_1^r\xi_2^s\xi_3^t=1 \right\}. \eeq The quotient
$C_d/G_d^{r,s,t}$ can be described algebraically
via projections \bea T_d^{r,s,t}: C_d &\lra &C_d^{r,s,t} \nn \\
(x,y) &\mapsto & (x^d, x^ry^s) =:(u,v), \eea which map $C_d$ into
the curves \beq C_d^{r,s,t} = \left\{v^d = u^r(1-u)^s\right\}.
\eeq

For prime degrees the abelian varieties $A_d^{r,s,t}$ can be
defined simply as the Jacobians $J(C_d^{r,s,t})$ of the
projections $C_d^{r,s,t}$. When $d$ has nontrivial divisors $m|d$,
this definition must be modified as follows. Consider the
projected Fermat curves \bea C_d &\lra & C_m \nn \\ (x,y) &\mapsto
&(\bx, \by):= \left(x^{\frac{d}{m}},y^{\frac{d}{m}}\right), \eea
whose Jacobians can be embedded as $e: J(C_m) \lra J(C_d)$.
Composing the projection $T_d^{r,s,t}$ as \beq J(C_m)
\stackrel{e}{\lra} J(C_d) \stackrel{T_d^{r,s,t}}{\lra}
J(C_d^{r,s,t}) \eeq for all proper divisors $m|d$ leads to a
collection of subvarieties $\cup_{m|d} T_d^{r,s,t}(e(J(C_m)))$.
The abelian variety of interest then is defined as \beq
A_d^{r,s,t} = J(C_d^{r,s,t})/\cup_{m|d}T_d^{r,s,t}(e(J(C_m))).
\eeq

The abelian varieties $A_d^{r,s,t}$ are not necessarily simple but
it can happen that they in turn can be factored. This question can
be analyzed via a criterion of Shimura-Taniyama, described in
\cite{st61}. Applied to the $A_d^{r,s,t}$ discussed here the
Shimura-Taniyama criterion involves computing for each set
$H_d^{r,s,t}$ defined as \beq \rmH_d^{r,s,t} := \left\{a\in
(\mathZ/d\mathZ)^{\times}~|~<ar>+<aks>+<a\ell t>=d\right\} \eeq
another set $W_d^{r,s,t}$ defined as \beq W_d^{r,s,t} = \left\{a
\in (\mathZ/d\mathZ)^{\times}~|~ aH_d^{r,s,t} = H_d^{r,s,t}
\right\}. \eeq If the order $|W_d^{r,s,t}|$ of $W_d^{r,s,t}$ is
unity then the abelian variety $A_d^{r,s,t}$ is simple, otherwise
it splits into $|W_d^{r,s,t}|$ factors \cite{kr78}.

We adapt this discussion to the weighted case. Denote the index
set of triples $(r,s,t)$ parametrizing the one-forms of the
weighted curves $C_d \in \IP_{(1,k,\ell}[d]$ again by $\cI$. The
cyclic group $(\mathZ/d\mathZ)^{\times}$ again acts on $\cI$ and
produces a set of orbits \beq \cO_{r,s,t} =  [(r,s,t)] \in
\cI/(\mathZ/d\mathZ)^{\times}.\eeq Each of these orbits leads to
an abelian variety $A_{[(r,s,t)]}$ of dimension \beq \rmdim
A_{[(r,s,t)]} = \frac{1}{2} \vphi\left(d_0\right), \eeq where
$\vphi$ is the Euler function $\vphi(n)=\#\{m~|~(m,n)=1\}$, and
complex multiplication with respect to the field $F_{[(r,s,t)]} =
\mathQ(\mu_{d_0})$, where $d_0=d/\rmgcd(r,ks,\ell t)$. This leads
to an isogeny \beq i: J(C_d) ~\lra ~\prod_{[(r,s,t)]\in
\cI/(\mathZ/d\mathZ)^{\times}} A_{[(r,s,t)]}.\eeq

 The complex multiplication type of the abelian factors
 $A_{r,s,t}$ of the Jacobian $J(C)$ can be identified with the set
 $H_d^{r,s,t}$ via a homomorphism from $\rmH_d^{r,s,t}$ to the
Galois group. More precisely, the CM type is determined by the
subgroup $G_d^{r,s,t}$ of the Galois group of the cyclotomic field
that is parametrized by $\rmH_d^{r,s,t}$ \beq
 G_d^{r,s,t} = \left\{\si_a \in \rmGal(\mathQ(\mu_{d_0})/\mathQ)~|~a
 \in \rmH_d^{r,s,t}\right\} \eeq
by considering \beq (F,\{\phi_a\}) = (\mathQ(\mu_{d_0}), \{\si_a
~|~ \si_a \in G_d^{r,s,t}\}).\eeq

\vskip .2truein

\section{The Fermat quintic threefold}

\subsection{CM type}

Consider the projective threefold embedded in projective $4-$space
and defined by \beq X_5 = \left\{(z_0:z_1:\dots :z_5)\in \IP_4~|~
z_0^5 + \cdots + z_4^5 =0 \right\}.\eeq We can split $d=3=1+2=r+s$
and apply the Shioda-Katsura construction to obtain the
decompositions \beq \rmH^3(X_5) \oplus \rmH^1(C_5)(1) \cong
\rmH^3(C_5\times S_5)^{\mu_5} \oplus \rmH^1(X_5^0 \times
C_d)(1)\eeq and \beq \rmH^2(S_5) \cong \rmH^2(C_5 \times
C_5)^{\mu_5} \oplus d(d-2)\rmH^2(\IP_1) \eeq in terms of the
cohomology groups of the Fermat curve \beq C_5 = \{x^5 + y^5 +
z^5=0\} ~\subset ~\IP_2\eeq  and the Fermat surface $S_5$.

From this we see that the basic building block of the cohomology
decomposition is given by the plane projective curve $C_5$ which
has genus $g(C_5) = 6$. The index set $\cI$ \bea
\cI &=& \{(1,1,3),(1,3,1),(3,1,1),(1,2,2),(2,1,2),(2,2,1); \nn \\
   & & ~~(2,4,4),(4,2,4),(4,4,2),(3,3,4),(3,4,3),(4,3,3)\} \nn
\eea parametrizes a basis of the first cohomology group of $C_5$,
which can be written as \beq \rmH^1_{\rmdR}(C_5) = \{
\om_{r,s,t}=x^{r-1}y^{s-5}dx~|~(r,s,t)\in \cI\}. \eeq The action
of $(\ZZ/5\ZZ)^{\times}$ leads to the orbits \bea
\cO_{1,1,3} &=& \{(1,1,3),(2,2,1),(3,3,4),(4,4,2)\} \nn \\
\cO_{1,3,1} &=& \{(1,3,1),(2,1,2),(3,4,3),(4,2,4)\} \nn \\
\cO_{3,1,1} &=& \{(3,1,1),(1,2,2),(4,3,3),(2,4,4)\}  \eea

Hence the Jacobian decomposes into a product of three abelian
varieties \bea
J(C_5) &=& \prod_{\cO_{r,s,t} \in \cI/(\ZZ/5\ZZ)^{\times}} A_{r,s,t} \nn \\
    & & \nn \\
  &=&  A_{1,1,3} \times A_{1,3,1} \times A_{3,1,1}, \eea
  each of dimension $\vphi(5)/2=2$, which arise from the
Jacobians of the genus two curves \bea
C_5^{1,1,3} &=& \left\{v^5 - u(1-u) = 0 \right\} \nn \\
C_5^{1,3,1} &=& \left\{v^5 - u(1-u)^3 = 0 \right\} \nn \\
C_5^{3,1,1} &=& \left\{v^5 - u^3(1-u) = 0 \right\}, \eea obtained
via the maps $T_5^{r,s,t}$.

In order to check the simplicity of the abelian factors we can use
the criterion of Shimura-Taniyama, described above. Computing the
sets $W_5^{r,s,t}$ for any of the triplets $(r,s,t)$ shows that
the order of these groups is unity, hence all three factors are in
fact simple.

For the complex multiplication type we find from \beq
\rmH_5^{1,1,3} = \{a\in
(\ZZ/5\ZZ)^{\times}=\{1,2,3,4\}~|~<a>+<a>+<3a>=5\}=\{1,2\} \eeq
that $G_5^{1,1,3}=\{\si_1,\si_2\}$ and therefore the complex
multiplication type of $A_{1,1,3}$ is given by \beq
(\mathQ(\mu_5), \{\vphi\} = \{\si_1,\si_2\}).\eeq The remaining
factors are described in the same way.

More explicitly, we can use the maps $T_5^{r,s,t}$ to express the
differentials of $C_d$ invariant under the action of $G_5^{r,s,t}$
in terms of the $(u,v)$ coordinates of $C_5^{r,s,t}$ and observe
their transformation behavior under the map \beq (u,v) \mapsto
(u,\xi_5v). \eeq

\subsection{Fusion field and quantum dimensions}

The field of complex multiplication derived for the quintic is
given by the cyclotomic field $\mathQ(\mu_5)$ and embedded in this
field is the real subfield $\mathQ(\sqrt{5})$, generated by the
elements $(\xi_5 + \xi_5^{-1})$. To compare this to the number
field determined by the string we briefly recall some facts about
the corresponding Gepner model \cite{g88, g87}.

The underlying exactly solvable model of the quintic threefold is
determined by the affine Kac-Moody algebra SU(2) at conformal
level $k=2$. The central charge $c(k) = 3k/(k+2)$ at level $k$
then leads to $c=9/5$, leading to a product of five models to make
a theory of total charge $c=9$. The physical spectrum of this
model is constructed from world sheet operators of the individual
SU(2) factors with the anomalous dimensions \beq \Delta^{(k)}_j =
\frac{j(j+2)}{4(k+2)},~~j=0,...,k, \eeq leading in the case $k=3$
to $\Delta^{(3)}_j \in \left\{0,\frac{3}{20}, \frac{2}{5},
\frac{3}{4}\right\}$.

These anomalous dimensions can be mapped into the quantum
dimensions $Q_{ij}$ via the Rogers dilogarithm \beq L(z) = Li_2(z)
+{\small \frac{1}{2}} \log(z)~\log(1-z), \eeq where $Li_2$ is
Euler's classical dilogarithm \beq Li_2(z) = \sum_{n\in \IN}
\frac{z^n}{n^2}, \eeq via the relation\cite{nrt92, k92, k94} \beq
\frac{1}{L(1)} \sum_{i=1}^k L\left(\frac{1}{Q_{ij}^2}\right) =
\frac{3k}{k+2} - 24 \Delta_j^{(k)} +6j \lleq{nrt} Here the
$Q_{ij}$ are defined as \beq Q_{ij} = \frac{S_{ij}}{S_{0j}}, \eeq
where the  \beq S_{ij} =
\sqrt{\frac{2}{k+2}}~~\sin\left(\frac{(i+1)(j+1)\pi}{k+2}\right),~~~~~0\leq
i,j \leq k \eeq describe the modular behavior of the SU(2) affine
characters \beq \chi_i\left(-\frac{1}{\tau}, \frac{u}{\tau}\right)
= e^{\pi i ku^2/2} \sum_j S_{ij} \chi_j(\tau, u). \eeq Applying
this map to the theory at conformal level three leads to the
quantum dimensions $Q_i = Q_{i0}$
 \beq Q_i\left({\rm SU(2)}_3\right)
  \in \left\{1,\frac{\tiny 1}{\tiny 2}(1+\sqrt{5})\right\}
  \subset ~\mathQ(\sqrt{5}).
\eeq

\vskip .2truein

\section{Appendix: Some number theory}

The ad\`elic language is a useful tool in the analysis of torsion
points of abelian varieties. The construction of this group
proceeds by first considering $\pfrak-$adic number fields and then
pasting these together in such a way that one obtains something
that is still manageable.

\subsection{$\pfrak-$adic numbers}

For rational primes $p$ $p-$adic number fields are constructed
much like the real number field $\mathR$ as completion of the
rational number field $\mathQ$ with respect to norms $|\cdot |_p$,
which are defined for each rational prime $p$ by noting that each
rational number $x\in \mathQ$ can be written as $x=\frac{m}{n}p^r$
for some $r\in \mathN$ and $m,n\in \mathZ$. The exponent $r$ is
called the order $\rmord_px=r$ of $x$ at $p$, and one writes
$|x|_p = p^{\rmord_px}$. In order to introduce a common language
one calls the norms for the reals and the complex numbers the
norms of the 'infinite primes' (for no particularly good reason).
More precisely, consider an algebraic number field $K$ with degree
$[K:\mathQ] = r_1+2r_2$, its real embeddings
$\{\si_1\}_{i=1,...,r_1}$, and its complex embeddings
$\{\si_{r_1+i}, \bsi_{r_1+i}\}_{i=1,...,r_2}$. Then for each real
$\si_i$ one defines a valuation on $K$ via \beq |x|_{\si_i} =
|x^{\si_i}|, \eeq where $x^{\si_i}$ denotes the action of $\si_i$
on $x\in K$. For the complex embeddings  define the valuation as
\beq |x|_{\si_{r_1+i}} = x^{\si_i} x^{\bsi_i}.\eeq These
valuations satisfy the archimedean inequality and are therefore
called archimedean. They define a metric with respect to which $K$
is a topological field, and one can consider the completions,
denoted by $K_{\infty,i}$, of $K$ with respect to these
valuations. This proceeds in complete analogy to the construction
of the real number field as the completion of the rational number
field with respect to the standard norm.

This construction can be generalized to other number fields via
valuations derived from the finite primes, i.e. prime ideals
$\pfrak \subset \cO_K \subset K$, where $\cO_K$ denotes the ring
of algebraic integers in $K$. Define for any $x\in K$ the
valuation $|\cdot|_{\pfrak}$ as \beq |x|_{\pfrak} = \rmN
\pfrak^{-\rmord_{\pfrak} x}, \eeq where $\rmN \pfrak=
\#(\cO_K/\pfrak)$ is the norm of the ideal $\pfrak$, and
$\rmord_{\pfrak} x$ is the power to which the ideal $\pfrak$
occurs in the factorization of the principal ideal $(x)$. Each
such valuation again defines a metric on $K$,
$d_{\pfrak}=|x-y|_{\pfrak}$, and one can again consider the
completion with respect to this metric, denoted by $K_{\pfrak}$.
The multiplicative subgroup is denoted by $K_{\pfrak}^{\times}$,
and the ring of integers in this field is defined by \beq
\cO_{\pfrak} = \left\{y\in K_{\pfrak}~|~|y|_{\pfrak} \leq
1\right\},\eeq while its group of units is given as \beq
\cO_{\pfrak}^{\times} = \left\{y\in K_{\pfrak}~|~ |y|_{\pfrak}
=1\right\}.\eeq

\subsection{Ad\`eles and id\`eles}

It is often useful in number theory to consider an embedding of a
number field of degree $n=r_1+2r_2$ into a vector space spanned by
its completions with respect to the standard metric ( \cite{rs01}
contains an application in a physical context). Given the
completions of $K$ with respect to finite and infinite primes, it
makes sense to try and put them all together into a single large
product $\prod_{\pfrak} K_{\pfrak}$ that combines all of them. It
turns out that this object is too large, and that it is more
useful to define the ring of ad\`eles $\mathA_K$ of the field $K$
to be the subset defined by a restricted product, consisting of
all elements $x=(x_{\pfrak})$ such that $x_{\pfrak}$ is a
$\pfrak-$adic integer for all but finitely many non-archimedean
valuations $|\cdot|_{\pfrak}$ \beq \mathA_K =
\left\{(x_{\pfrak})~|~ x_{\pfrak} \in K_{\pfrak} ~\forall \pfrak,
x \in \cO_{\pfrak}~{\rm for~all~but~finitely~many~}\pfrak
\right\}. \eeq

Put somewhat differently, consider a set $S$ of primes that
contains the infinite ones. Then we can construct the ring \beq
\mathA_K^S = \prod_{\pfrak \notin S} \cO_{\pfrak} \prod_{\qfrak
\in S} K_{\qfrak}, \eeq where the multiplication is understood to
be component wise. For $S\subset S'$ there is an injection
$\mathA_S \lra \mathA_{S'}$ and one can define \beq \mathA_K =
\lim_S \mathA_K^S.\eeq

Embedded within the ad\`eles is the multiplicative subgroup
$\mathA_K^{\times}$ called the id\`eles. These were originally
introduced by Chevalley \cite{c36-40}, independently of the later
notion of ad\`eles \cite{aw45}, and can be described in a way
completely analogous to the ad\`eles as \beq \mathA_K^{\times} =
\left\{(x_{\pfrak})~|~ x_{\pfrak} \in K_{\pfrak}^{\times} ~\forall
\pfrak, ~x \in \cO_{\pfrak}^{\times}~{\rm
for~all~but~finitely~many~}\pfrak \right\}. \eeq Addition and
multiplication in the ad\`eles and id\`eles are understood to be
component wise.

If  $K/F$ is a finite extension then one can define an id\`elic
norm map \beq \rmN^K_F: \mathA_K^{\times} \lra \mathA_F^{\times}
\eeq which is determined by specifying what the $v-$th component
is of the image id\`ele, where $v$ describes either the finite of
infinite primes. For $x\in \mathA_K^{\times}$ one sets \beq \left(
\rmN^K_F x\right)_v = \prod_{w|v} \rmN^{K_w}_{F_v} x_w,\eeq where
the product is over all $w$ which contain $v$, and $K_w, F_v$
denote the completions of $K$ and $F$ with respect to the norms
defined by $w$ and $v$ respectively.

\subsection{Class field theory}

Class field theory deals with the behavior of primes in number
fields $K$ when $K$ is embedded in finite extensions $L/K$. For
simple fields such as quadratic extensions of the rationals this
behavior is controlled by the Legendre symbol. For more
complicated fields the Artin symbol provides an appropriate
generalization.

Consider a finite number field extension $L/K$ that is Galois and
an ideal $\Pfrak \subset \cO_L$ in the ring of integers $\cO_L$ of
$L$ that divides the prime ideal $\pfrak \subset \cO_K$. The Artin
symbol is then defined as the element $\si_{\Pfrak}$ of the Galois
group $\rmGal(L/K)$, often denoted by $(\Pfrak, L/K)$, or
$\left(\frac{L/K}{\Pfrak}\right)$, for which \beq \si_{\pfrak}(x)
= x^{\rmN\pfrak}~\rmmod ~\Pfrak.\eeq It is a multiplicative
function on the primes \beq \left(\frac{L/K}{\cdot}\right): \cI(L)
\lra \rmGal(L/K) \eeq and therefore can be defined for any ideal
$\Afrak \subset \cO_L$ via the prime decomposition $\Afrak = \prod
\Pfrak_i$ as \beq \left(\frac{L/K}{\Afrak}\right) = \prod_i
\left(\frac{L/K}{\Pfrak_i}\right). \eeq When the extension $L/K$
is abelian the Artin symbol is independent of the prime $\Pfrak$
chosen above $\pfrak$ and depends only on $\pfrak$.

The id\`elic formulation of class field theory aims at an
efficient discussion of all abelian extensions at the same time.

{\bf Theorem.}~ {\it Let $K$ be a number field and $K_{\rmab}$ be
the maximal abelian extension of $K$. There exists a unique
continuous homomorphism, called the reciprocity map} \bea
\mathA_K^{\times} &\lra & \rmGal(K_{\rmab}/K) \nn
\\ x &\mapsto & [x,K],\eea {\it with the property that if
$L/K$ is a finite abelian extension, $x\in \mathA_K^{\times}$ an
id\`ele whose ideal $(x)$ is not divisible by any primes that
ramify in $L$, then \hfill \break (1)~$[x,K]_L = ((x),L/K)$ is the
Artin symbol of the ideal $(x)$. \hfill \break 2) The reciprocity
map is compatible with the norm map: if $L$ is a finite abelian
extension of $K$ then} \beq [x,L]|_{K_{\rmab}} =
\left[\rmN^L_Kx,K\right],~~~\forall x\in \mathA_L^{\times}.\eeq

\vskip .2truein

{\bf Acknowledgement} \hfill \break It is a pleasure to thank Jack
Morse for correspondence. M.L. and R.S. are grateful to the Kavli
Institute for Theoretical Physics, Santa Barbara, for hospitality
and support through KITP Scholarships during the course of part of
this work. This research was supported in part by the National
Science Foundation under grant No. PHY99-07949.


\vskip .2truein

\begin{thebibliography}{9}
\bibitem{w49} A. Weil, {\it Number of solutions of equations in finite
fields}, Bull. Amer. Math. Soc. {\bf 55} (1949) 497
\bibitem{w52} A. Weil, {\it Jacobi sums as "Gr\"o\ss encharaktere"}, Trans.
Amer. Math. Soc. {\bf 73} (1952) 487
\bibitem{y01} N. Yui, {\it The Arithmetic of certain Calabi-Yau varieties
over number fields}, B.B. Gordon, J.D. Lewis, S. M\"uller-Stach,
S. Saito and N. Yui (eds), {\sc The Arithmetic and Geometry of
Algebraic Cycles}, Kluwer, 2000
\bibitem{rs01} R. Schimmrigk, {\it Arithmetic Geometry of
Calabi-Yau Varieties and Rational Conformal Field Theory}, J.
Geom. Phys. {\bf 44} (2003) 555 - 569,  arXiv: hep-th/0111226
\bibitem{su02} R. Schimmrigk and S. Underwood, {\it The
Shimura-Taniyama Conjecture and Conformal Field Theory}, J. Geom.
Phys. {\bf 48} (2003) 169 - 189, arXiv: hep-th/0211284
\bibitem{bcdt01} C. Breuil, B. Conrad, F. Diamond and R. Taylor, {\it On the
modularity of elliptic curves over $\mathbb{Q}$ or Wild 3-adic
exercises}, J. Amer. Math. Soc. {\bf 14} (2001) 843 - 939
\bibitem{w95} A. Wiles, {\it Modular elliptic curves and Fermat's Last
Theorem}, Ann. Math. {\bf 141} (1995) 443 - 551; \\
 R. Taylor and A. Wiles, {\it Ring-theoretic properties of
certain Hecke algebras}, Ann. of Math. {\bf 141} (1995) 553 - 572
\bibitem{l79}  R. Langlands, {\it Automorphic representations, Shimura varieties,
and motives. Ein M\"archen}, Proc. Symp. Pure Math. XXXIII, Vol.
2, 1979
\bibitem{d50s} M. Deuring, {\it Die Zetafunktion einer algbebraischen
Kurve vom Geschlecht Eins, I - IV}, Nachr. Akad. Wiss.
G\"ottingen, Math. - Phys - Chem. Abt, 1953, S. 88-94; 1955, S. 13
-42; 1956, S. 37 - 76;  1957, S. 55 - 80
\bibitem{gv02} S. Gukov and C. Vafa, {\it Rational Conformal Field Theories and
Complex Multiplication}, arXiv: hep-th/0203213
\bibitem{m69} D. Mumford, {\it A Note on Shimura's Paper {\sc
Discontinuous Groups and Abelian Varieties}}, Math. Ann. {\bf 181}
(1969) 345
\bibitem{pss73} I. Pjateckii and I.R. Shafarevich, {\it The
Arithmetic of K3 Surfaces}, Proc. Steklov Inst. Math. {\bf 132}
(1973) 45
\bibitem{b92}  C. Borcea, {\it Calabi-Yau Threefolds and Complex
Multiplication}, in {\sc Essays on Mirror Manifolds}, ed. S.-T.
Yau, International Press, 1992
\bibitem{y03} N. Yui, {\it L-series of Calabi-Yau Orbifolds of CM
Type}, preprint
\bibitem{lps03} M. Lynker, V. Periwal and R. Schimmrigk, {\it Complex
Multiplication Symmetry of Black Hole Attractors}, Nucl. Phys.
{\bf B667} (2003) 484 - 504, arXiv: hep-th/0303111
\bibitem{m98}  G. Moore, {\it Arithmetic and Attractors},
arXiv: hep-th/9807057
\\ {\it Attractors and Arithmetic}, arXiv: hep-th/9807086
\bibitem{d74} P.Deligne,  {\it La conjecture  de Weil I},
Publ.Math. IHES {\bf 43}(1974)273
\bibitem{ea24} E. Artin, {\it Quadratische K\"orpher in Gebiete
der h\"oheren Kongruenzen I \& II}, Math. Zeit. {\bf 19} (1924)
153-246
\bibitem{fks31} F. K. Schmidt, {\it Analytische Zahlentheorie in
K\"orpern der Charakteristik $p$}, Math. Zeit. {\bf 33} (1931 1-32
\bibitem{hh34} H. Hasse, {\it \"Uber die Kongruenzzetafunktionen.
Unter Benutzung von Mitteilungen von Prof. Dr. F.K. Schmidt und
Prof. Dr. E. Artin}, S. Ber. Preu\ss. Akad. Wiss. H. {\bf 17}
(1934) 250 - 263
\bibitem{bd60} B.Dwork, {\it On the rationality of the zeta
function of an algebraic variety}, Amer.J.Math. {\bf 82}(1960)631
\bibitem{h18}  E. Hecke, {\it Eine neue Art von Zetafunktionen und
ihre Beziehungen zur Verteilung der Primzahlen}, Math. Z. {\bf 1}(1918)357;\\
{\it Eine neue Art von Zetafunktionen und ihre Beziehungen zur
Verteilung der Primzahlen. Zweite Mitteilung}, Math. Z. {\bf 6}
(1920)11
\bibitem{d77} P.Deligne, {\it Cohomologie \'etale}, Springer Verlag LNM 569, 1977
\bibitem{js97}  J. H. Silverman, {\it A Survey of the Arithmetic Theory
of Elliptic Curves}, in {\sc Modular Forms and Fermat's Last
Theorem}, Springer 1997
\bibitem{js94}  J. Silverman, {\it Advanced Topics in the
Arithmetic of Elliptic Curves}, Springer, 1994
\bibitem{gs71a} G. Shimura, {\it Introduction to Arithmetic Theory
of Automorphic Functions}, Princeton University Press, 1971
\bibitem{s66} J.-P. Serre, {\it Complex multiplication}, in {\sc
Seminar on Complex Multiplication}, eds. A. Borel, S. Chowla, C.S.
Herz, K. Iwasawa, J.-P. Serre, LNM 21, Springer Verlag, 1966
\bibitem{c36-40} C. Chevalley, {\it G\'en\'eralisation de la
th\'eorie du corps de classes pour les extensions infinies}, J.
Math. Pure App. {\bf 15} (1936) 359 - 371; \\
{\it La th\'eorie du corps de classes}, Ann. Math. {\bf 41} (1940)
394 - 418
\bibitem{aw45} E. Artin and G. Whaples, {\it Axiomatic
characterization of fields by the product formula for valuations},
Bull. Amer. Math. Soc. {\bf 51} (1945) 469 - 492
\bibitem{h27} H. Hasse, {\it Neue Begr\"undung der komplex Multiplikation},
 J. Reine Angew. Math. {\bf 157} (1927) 115 - 137; {\bf 165} (1931) 64 - 88
\bibitem{n90} W. Narkiewicz, {\it Elementary and analytic theory
of algebraic numbers}, Springer 1990
\bibitem{d49} M. Deuring, {\it Algebraische Begr\"undung der
komplexen Multiplikation}, Abh. Math. Sem. Hamburg {\bf 16} (1949
32 - 47
\bibitem{t57} Y. Taniyama, {\it L-functions of number
fields and zeta functions of abelian varieties}, J. Math. Soc.
Japan {\bf 9} (1957) 330 - 366
\bibitem{st61} G. Shimura and Y. Taniyama, {\it Complex
Multiplication of Abelian Varieties and its Application to Number
Theory}, 1961
\bibitem{st68} J.-P. Serre and J. Tate, {\it Good Reduction of
Abelian Varities}, Ann. Math. {\bf 88} (1968) 492 - 517
\bibitem{gs71b} G. Shimura, {\it On the zeta-function of an
abelian variety with complex multiplication}, Ann. Math. {\bf 94}
(1971) 504 - 553
\bibitem{sk79} T. Shioda and T. Katsura, {\it On Fermat
varieties}, Tohoku Math. J. {\bf 31} (1979)97-115
\bibitem{d82} P. Deligne, {\it Hodge cycles on abelian varitites},
in {\sc Hodge Cycles, Motives, and Shimura Varieties}, Eds. P.
Deligen, J.S. Milne, A. Ogus and K.-y. Shen, LNM 900, Springer
1982
\bibitem{cls90} P. Candelas, M. Lynker, and R. Schimmrigk, {\it
Calabi-Yau Manifolds in Weighted $\IP_4$}, Nucl. Phys. {\bf B341}
(1990) 383 - 402, KEK archive 199001117
\bibitem{s87} R. Schimmrigk, {\it A New Construction of a
Three-generation Calabi-Yau Manifold}, Phys. Lett. {\bf B193}
(1987) 175, KEK archive 198706386; \\
{\it Heterotic RG Flow Fixed Points with Non-Diagonal Affine
Invariants}, Phys. Lett. {\bf B229} (1989) 227, KEK archive
198905565
\bibitem{f61} D. K. Faddeev, {\it On the Divisor Groups of some
Algebraic Curves}, Dokl. Tom {\bf 136} (1961) 296-298 [Sov. Math
{\bf 2}(1961) 67-69] \\ {\it Invariants of Divisor Classes for the
Curves $y^{\ell} = x^k(1-x)$ in an $\ell$-adic Cyclotomic Field},
Trudy Mat. Inst. Steklova {\bf 64} (1961) 284 - 293 (In Russian)
\bibitem{w76} A. Weil, {\it Sur les P\'eriodes des Int\'egrales
  Abeliennes}, Comm. Pure Appl. Math {\bf 29} (1976) 813
\bibitem{g78} B. H. Gross, {\it On the Periods of Abelian Integrals
and a Formula of Chowla and Selberg}, Inv. math. {\bf 45} (1978)
193
\bibitem{r78} D. E. Rohrlich, {\it The Periods of the Fermat Curve},
Appendix to Gross 1978, Inv. math. {\bf 45} (1978) 208
\bibitem{kr78} N. Koblitz and D. E. Rohrlich, {\it Simple Factors
in the Jacobian of a Fermat Curve}, Can. J. Math. 30 (1978) 1183 -
1205
\bibitem{g88} D. Gepner, {\it Spacetime supersymmetry in
compactified string theory and superconformal models}, Nucl. Phys.
{\bf B296} (1988) 757
\bibitem{g87} D. Gepner, {\it Exactly solvable string
compactification and manifolds of SU($N$) holonomy}, Phys. Lett.
{\bf B199} (1987) 380
\bibitem{nrt92} W. Nahm, A. Recknagel and M. Terhoeven, {\it
Dilogarithm Identities in Conformal Field Theory}, Mod. Phys.
Letts. {\bf A8} (1993) 1835 - 1848, arXiv: hep-th/9211034
\bibitem{k92} A. N. Kirillov, {\it Dilogarithm Identities,
Partitions, and Conformal Field Theory}, arXiv: hep-th/9212150
\bibitem{k94} A.Kirillov, {\it Dilogarithm Identities}, Progr. Theor. Phys.
Suppl {\bf 118} (1995) 61, arXiv: hep-th/9408113
\end{thebibliography}
\end{document}